\documentclass{article}

\usepackage[utf8]{inputenc}
\usepackage[T1]{fontenc}
\usepackage{amsmath}
\usepackage{amsthm}
\usepackage{amssymb}
\usepackage{graphicx}
\usepackage{physics}
\usepackage{todonotes}
\usepackage{pgf,tikz,pgfplots}
\usepackage[english]{babel}
\usepackage[margin=1.1in]{geometry}
\usepackage{authblk}
\pgfplotsset{compat=1.16}
\usepackage{tabularx}
\usepackage{siunitx}
\usepackage{caption}
\usepackage{subcaption}
\usepackage{url}

\usepackage[normalem]{ulem}

\newcounter{protocol}

\begin{document}
\title{Connecting Quantum Cities: Simulation of a Satellite-Based Quantum Network}
\author[1,2,3]{Raja Yehia}
\author[1]{Matteo Schiavon}
\author[1,4]{Valentina Marulanda Acosta}
\author[5,6]{Tim Coopmans}
\author[2]{Iordanis Kerenidis}
\author[5,7]{David Elkouss}
\author[1]{Eleni Diamanti}

\affil[1]{Sorbonne Université, CNRS, LIP6, F-75005 Paris, France}
\affil[2]{Université de Paris, CNRS, IRIF, F-75013 Paris, France}
\affil[3]{ICFO - Institut de Ciencies Fotoniques, The Barcelona Institute of Science and Technology}
\affil[4]{DOTA, ONERA, Université Paris Saclay, F-92322 Châtillon, France}
\affil[5]{QuTech, Delft University of Technology, Lorentzweg 1, 2628 CJ Delft, The Netherlands}
\affil[6]{Leiden Institute of Advanced Computer Science, Leiden University, Niels Bohrweg 1, 2333 CA Leiden, The Netherlands}
\affil[7]{Networked Quantum Devices Unit, Okinawa Institute of Science and Technology Graduate University, Okinawa, Japan}

\date{\today}

\maketitle

\begin{abstract}
We present and analyse an architecture for a European-scale quantum network using satellite links to connect Quantum Cities, which are metropolitan quantum networks with minimal hardware requirements for the end users. Using NetSquid, a quantum network simulation tool based on discrete events, we assess and benchmark the performance of such a network linking distant locations in Europe in terms of quantum key distribution rates, considering realistic parameters for currently available or near-term technology. Our results highlight the key parameters and the limits of current satellite quantum communication links and can be used to assist the design of future missions. We also discuss the possibility of using high-altitude balloons as an alternative to satellites.

\end{abstract}

\section{Introduction}

The promises of the Quantum Internet, a global network of quantum devices connected through quantum channels~\cite{QIavision}, are numerous. Such an infrastructure can enhance the performance of current networks by increasing their security or communication efficiency~\cite{SecurityQKD,Leaderelection,SecretSharing,RefCommComplexity}, allows for new functionalities such as, for instance, secure delegated computing~\cite{delegated1,DelegatedQC}, and opens the way to distributed quantum computing and sensing~\cite{DANOS200773,shettell2022private}. Quantum network protocols between two or multiple parties~\cite{MEVresistant,Anonymity,fedeVoting,ConferenceKeyAgreement} executed over the Internet require quantum states to be transferred across the infrastructure, typically in the form of shared entanglement. This is achieved by routing quantum states through the network via teleportation, which consumes the entanglement of so-called Bell (or EPR) pairs~\cite{Meignant_2019,Bughalo_2023}. The ability to efficiently share entanglement between any number of remote nodes of the network is therefore a fundamental resource for Quantum Internet applications and currently the topic of intense efforts aiming at optimal network architecture designs~\cite{QIRG}.\\

While increasingly advanced quantum networks are being deployed~\cite{BristolQCity,ChinaQKDNetwork}, it becomes crucial to consider more global strategies for connecting metropolitan or regional scale networks to reach continental scales and beyond. Carefully developed simulation tools can be extremely helpful to this end, and can inform the design of both near and longer term quantum networks, in particular at the present stage of small scale deployment, where it is desirable to avoid wasting resources on ultimately unscalable architectures.\\

One of the main obstacles towards building an international quantum network is long distance communication. After a few tens of kilometers, photon loss in fiber becomes predominant and prevents practical applications. Well known  bounds~\cite{PLOB,Takeoka_2014} give fundamental limits on quantum communication over long distance in a fiber. Quantum repeaters~\cite{Repeater1,Repeater2, sangouard2011quantum, azuma2022quantum}, the quantum analogues of signal amplifiers in classical networks, will provide a solution to this problem. However, despite significant experimental advances in the recent past~\cite{Bhaskar_2020,Repeater3, azuma2022quantum,PompiliScience2021,Lago_Rivera_2021}, they do not have yet the necessary technological maturity~\cite{repeaterNV,Ruf_2021,avis2022requirements}. Over the last decade, free space communication and more particularly satellite communication has received more and more attention as a complementary solution to overcome this issue. Feasibility studies~\cite{Bonato2009,Bourgoin2013}, together with experimental implementations of ground-to-ground~\cite{SchmittManderbach2007}, airplane-to-ground~\cite{Nauerth2013} or balloon-to-ground~\cite{Wang2013} free-space channels, have led the way to some experimental studies of satellite-to-ground quantum communication exploiting a simulated quantum source on satellite~\cite{Vallone2015,Gnthner2017}. A crucial milestone in the race for satellite communication was the launch of the Chinese satellite Micius~\cite{Lu2022}, which has allowed for the first demonstration of prepare-and-measure~\cite{Liao2017} and entanglement-based~\cite{Yin2020} quantum key distribution and of other quantum protocols, such as quantum teleportation~\cite{Ren2017}, using a satellite terminal. In satellite communication, the main source of loss is given by diffraction, which scales quadratically with the distance, in stark contrast with the exponential loss in glass fiber.  Satellite quantum communication thus appears as a particularly appealing choice for interconnecting the metropolitan quantum networks under deployment, as demonstrated by the QKD network deployed in China, which extends over more than 4000 km~\cite{ChinaQKDNetwork}.\\

Considering that the use of satellite links will play a central role in the implementation of very large scale quantum networks, the design of a complete network infrastructure requires a way to simulate such links, including in particular atmospheric perturbations. Previous works have mostly focused on integrated models of a single passage of a satellite above two ground stations, where the atmospheric effects are modeled in an aggregated way through an average transmissivity, which moreover only takes into account attenuation and diffraction losses~\cite{Liorni2021,Khatri2021}, or  also the beam-wandering effects due to atmospheric turbulence and pointing error~\cite{Boone2015}. All these works consider a downlink channel, with the source on the satellite and the detectors in the ground station. This is justified by the fact that most of the travelling of the photons happens in space, where there is no atmosphere and the only loss and noise are due to diffraction; the fact that the atmosphere affects the beam only in the last part of the channel gives a lower beam-wandering effect than in the uplink case, where these effects are at the beginning of the propagation.
For this reason, in this work we will also concentrate on the downlink scenario.\\

However, contrary to previous works, we consider an atmospheric model that provides, for each photon traveling through the channel, the instantaneous value of the transmissivity. Considering the effect of transmission at each point of the satellite is relevant for the performance of satellite quantum communication because the distance between the satellite and the ground stations is not constant and also due to the limited time window during which a satellite is in view of a ground station.
In this work, we quantify this effect by simulating each photon, emitted from a moving satellite, individually. \\

Our simulation tool is based on NetSquid~\cite{coopmans2021netsquid, Netsquid}, a quantum network simulator with discrete events developed at QuTech in the Netherlands. We have previously developed using NetSquid a set of subroutines allowing the simulation of a metropolitan terrestrial quantum network, with which we studied the feasibility of different protocols~\cite{QCity}. Here, we add satellite nodes, noise models relevant for satellite to ground links and routines to simulate satellite orbits. The code used in our work is available on GitHub \cite{github,githubMatteo}.\\

\begin{figure}[!ht]
    \centering
    \includegraphics[width=15cm]{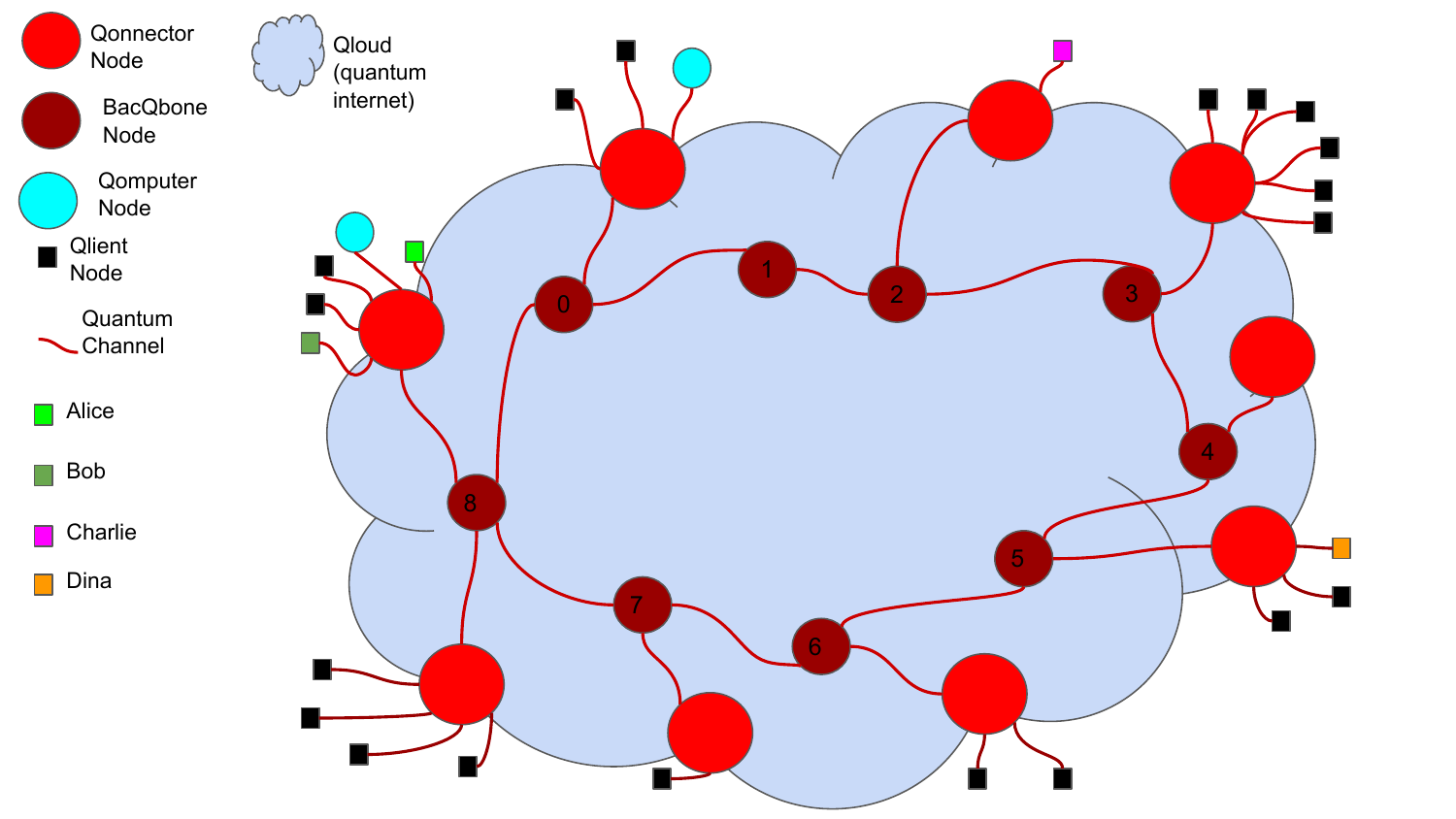}
    \caption{Schematic of a Qloud: Quantum Cities connected through a backbone network of satellites (bacQbone nodes). A Quantum City is formed by a powerful central node (Qonnector) used as a server allowing end users (Qlients) to enjoy quantum-enhanced functionalities. The end users can also be powerful quantum machines (Qomputer nodes). }
    \label{fig:Qloud}
\end{figure}

We use these tools to investigate the feasibility of a satellite-based  network architecture in a realistic  context within Europe. We show our envisioned network architecture that we call the Qloud in Fig~\ref{fig:Qloud}. In this architecture, users with minimal quantum technology abilities, called Qlients, are connected to central nodes called Qonnectors, together forming a star network which we refer to as Quantum City. In turn, the Qonnectors are connected to other Qonnectors through a network of bacQbone nodes, thereby enabling quantum communication between any two users in the network. This setup is very general to allow putting as few restrictions as possible on potential future extensions or applications, while minimizing the hardware requirements on the end users. This is key for practical implementations. In this work, we investigate quantum communication between Quantum Cities based on satellite links acting as bacQbone nodes between Qonnectors.

We investigate a small Qloud with simulations using real satellite data that show the feasibility of quantum key distribution (QKD) in this network. QKD is one of the most studied applications of quantum communication and we use it here as a performance benchmark for the network under study. We first examine the critical parameters of satellite quantum communication by looking separately at their effect on the key rate in a simple BB84 downlink scenario. Then, we analyse and discuss the practical relevance of such links in different realistic network settings. More precisely this work is organized as follows. After briefly recalling the characteristics of a Quantum City in Sec.~\ref{sec:ArchDesc}, we detail our model for simulating satellite quantum communication to connect Quantum Cities. Then in Sec.~\ref{sec:Result} we focus on a simple downlink scenario and employ our simulation code for an exploratory case: first, we determine a suitable satellite which we choose for later investigations, and then we characterise the influence of the various parameters in our model. In Sec.~\ref{sec:QKD}, we embed this study in the context of quantum networks and obtain the achievable raw QKD rate between two distant users in different settings. Finally, in Sec.~\ref{sec:discussion}, we use these results to discuss the perspectives of satellite quantum communication in the near future as well as some alternatives such as the use of high-altitude balloons, which has seen a rising interest in the last few years~\cite{Moll2019,Vu2020,Chu2021}.

\section{Architecture description}
\label{sec:ArchDesc}
\subsection{Quantum City}
In previous work~\cite{QCity}, we introduced the Quantum City architecture for metropolitan photonic quantum networks that minimizes the hardware necessary for the users while still enhancing current classical networks. In this section, we recall the main features of this architecture.\\

\begin{figure}[!ht]
    \centering
    \includegraphics[width = 10cm]{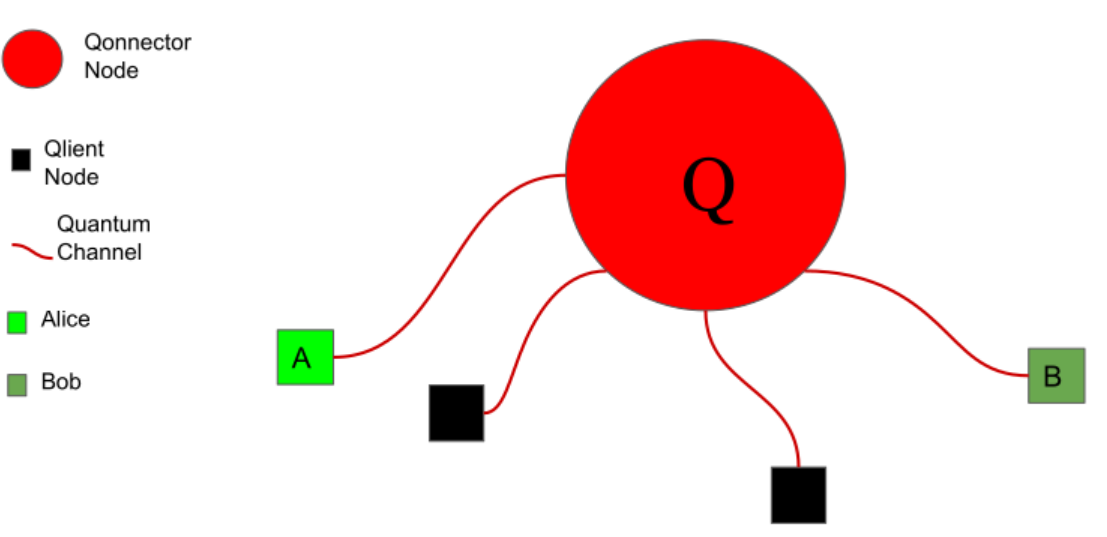}
    \caption{The Quantum City topology. It is a star-type photonic quantum network with a special node in the middle, the Qonnector, that has high quantum capabilities. Each end node (Qlient, such as Alice or Bob) has limited quantum hardware and is connected to the Qonnector through an optical fiber. }
    \label{fig:ArQitecture}
\end{figure}

Our model considers a star topology with a central node, that we call Qonnector, linked to every user, that we call a Qlient, through optical fibers (see Fig.~\ref{fig:ArQitecture}). This allows for centralized routing procedures and asymmetrical distribution of hardware between a powerful Qonnector and limited Qlients. This corresponds to expected intermediate-term development of quantum hardware and its availability to end users. Below we describe precisely the abilities of a Qonnector and a Qlient node in our model.

\textbf{Qonnector} nodes are at the core of the Quantum City model. They abstract servers providing quantum services to end users. They are powerful nodes that can create, send and receive any state, and that are connected both classically and quantumly to a certain number of parties. We suppose a Qonnector has currently available state-of-the-art photonic capabilities, namely that it can create and manipulate single-qubit states as well as two and multiparty entangled states such as Bell pairs and GHZ states. We also suppose that Qonnectors have access to measurement devices allowing them to receive and measure qubit states and to perform probabilistic Bell state measurements on two qubits arriving simultaneously. Finally we suppose that Qonnectors have the ability to route photonic states arriving from a Qlient to another one and that they have access to usual classical computing power and classical Internet. In this work, Qonnectors also represent ground stations for satellite communication. We suppose that they are equipped with telescopes and measurement stations allowing them to receive and process photonic quantum states coming from a satellite. We also suppose that the Qonnector is able to couple the states coming from a satellite into a fiber and route them to any Qlient.\\

\textbf{Qlient} nodes represent the end users connected to the quantum internet. They are abstractions of private users that hold near-term photonic quantum communication devices. They are classically connected to the rest of the network through the classical Internet and have usual classical computing power. We suppose that they have very limited quantum hardware capabilities, namely they manipulate one qubit at a time. More precisely they can create, send and manipulate any single-qubit photonic state as well as receive and measure it. Industrial-grade devices offering these capabilities are already available today or will become in the near future, and can be expected to become more suitable for wider use in the following years, thanks to advances for instance in photonic integration~\cite{PhotonicIntegrationRoadmap}.\\

For more details on these nodes and how they are modeled in our simulations, please see~\cite{QCity}.

\subsection{Connecting Quantum Cities with satellites}
\label{subsec:SatModel}

Let us strat our analysis by detailing our model for the noise in satellite to ground quantum communication. Since atmospheric turbulence is a very complex phenomenon, the realisation of an adequate model requires a choice of the effects that most severely affect the transmission. Here, we focus on atmospheric absorption and beam wandering effects.\\

In order to describe the impact of atmospheric propagation on the signal, we use a computer code called Lowtran~\cite{Kneizys1988}, which is a Fortran code developed for the calculation of the transmittance and background radiance of the atmosphere. It is based on an empirical prediction scheme derived from lab measurements and provides a reasonably accurate estimation of atmospheric effects over a broad spectral interval ($\sim$ 0.25 to 28.5 $\mu$m). We briefly describe here the parameters that are included in the model of the atmosphere in Lowtran and refer to~\cite{Kneizys1988} for details. The atmosphere is represented by 33 horizontal layers between sea level and an altitude of 100 km. The total transmittance is computed as the product of different atmospheric effects, namely continuum absorption, aerosol extinction, molecular scattering and molecular absorption, the latter of which includes the influence of water vapor, ozone, nitric acid and other uniformly mixed gases. The program contains a few different representative atmospheric models based on geographical-seasonal characteristics (such as for tropical or mid-latitude environments) that encompass the variations of pressure, temperature, water vapor and ozone with altitude. It also accounts for several aerosol models that describe particular meteorological ranges such as an urban environment, a less severe rural setting or a more wind and humidity dependent maritime navy situation. Last, a few different visual ranges corresponding to different aerosol density models are considered as well~\cite{kneizys1978atmospheric}.\\

In order to account for beam wandering, we used the model proposed in~\cite{SatelliteModel}, which presents a rigorous treatment of beam wandering effects, one of the leading causes of losses in a free-space channel. Its main advantage lies in an analytical formulation of the probability distribution of the transmission coefficient (PDTC), a feature used to provide a computationally efficient software implementation of the model. The model has also been applied recently to the analysis of continuous-variable quantum key distribution over a satellite-to-ground channel ~\cite{LuisVictorFeasability}. While a subsequent, more complete model of atmospheric propagation for the satellite-to-ground case is described in ~\cite{SatelliteModelComplex}, its much higher complexity makes it challenging to embed this model in NetSquid.\\

Beam wandering effects come from two main sources, namely the turbulence and the jitter due to the pointing error of the transmitter. The effects of beam wandering due to turbulence depend mostly on the size of the beam at the beginning of the propagation in the atmosphere and are determined by the refractive index structure constant $C_n^2$ which we will consider as fixed throughout the propagation. The satellite pointing jitter is, in turn, characterized by the standard deviation of the pointing error $\theta_p$. The parameters that are necessary to physically describe the channel are the size of the transmitting and receiving stations and the properties of the atmosphere and the pointing system. The main characteristic of the receiving station is the radius of the receiving telescope, which determines the proportion of the transmitted light that can be collected by the receiver. \\

The current code considers two possible configurations for this channel, namely the ground-to-ground free-space one (class FreeSpaceLossModel) and the satellite-to-ground one (class FixedSatelliteLossModel). The first one considers a horizontal channel, meaning that the entire propagation path is within the atmosphere and therefore will be affected by it. This can either be the case for direct communication between two ground stations or for a link between two drones or high altitude balloons carrying telescopes. In both cases, the $C_n^2$ value is indeed constant and will depend on the altitude of the link. For this configuration, the transmitter is characterized by the beam waist $\omega_0$.\\

The second configuration considers a slant propagation path, only the last 10 km of which will be affected by the atmosphere. For this kind of links, a constant $C_n^2$ is more of an approximation since in reality it varies throughout the propagation path as it is a highly altitude dependent parameter. For this configuration, the transmitter is characterized by the divergence angle $\theta_d$, a value related to the previously mentioned beam waist as follows: $\theta_d = \lambda/(\pi \omega_0)$.\\

As for the pointing error of the satellite, assuming that the position of the center of the transmitted beam with respect to the receiving aperture follows a normal distribution and is centered around the midpoint of the aperture, the PDTC will follow a log-negative Weibull distribution. The incidence of turbulence on beam wandering is less important for the satellite-to-ground case, becoming negligible in front of the beam wandering effects due to the pointing error $\theta_p$.\\

The model assumes that each qubit is affected by the PDTC independently from the other qubits of the transmission. Despite being unrealistic since it neglects the dynamics of the atmosphere, which is considerably slower than the typical time delay between two qubits, it allows to provide a good insight of the average properties of the channel. In addition to this, the satellite-to-ground channel model assumes a fixed position for the satellite. This allows to give a first estimate of the performance of the channel when the satellite is on a given position in the sky, but it lacks the ability to provide information about a long-time operation on the channel. We take this into account externally by discretizing the orbit in 10 second intervals for which the satellite is considered as fixed and we then make a separate simulation for each trajectory.\\

In the following, we will use the satellite as a bacQbone node to connect two Quantum cities (see Fig.~\ref{fig:Qloud}). We will suppose it is able to create and send single-qubit, BB84-type states with probability $p_\text{qubit}$ at a time rate $f_\text{qubit}$ as well as EPR states with probability $p_\text{EPR}$ and a time rate $f_\text{EPR}$ to two ground stations. We will use the terms `raw rate' or `QKD rate', or simply rate, over a network for the fraction $\frac{n_{\textnormal{arrived}}}{n_{\textnormal{sent}}}$, where $n_{\textnormal{arrived}}$ is the number of quantum states (either single-qubit photon states or EPR pairs) that actually arrived at the desired receiving nodes and $n_{\textnormal{sent}}$ is the number of quantum states that are sent over the network.

\section{Simulation results}
\label{sec:Result}

\subsection{Setting and parameters}
\label{sec:param}
\begin{figure}[!ht]
    \centering
    \includegraphics[width=\textwidth]{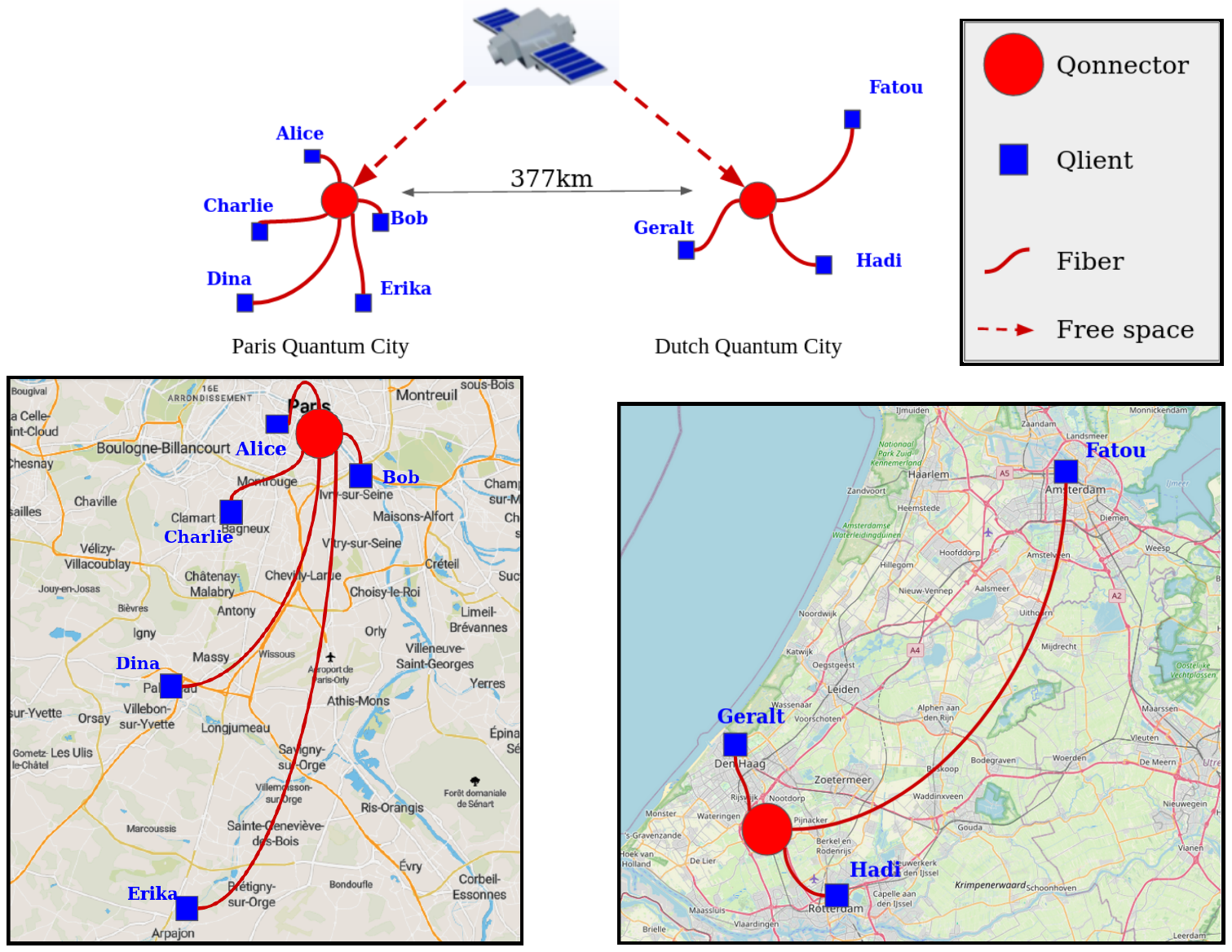}
    \caption{A satellite connecting Paris and Dutch Quantum Cities with downlinks only. Quantum City of Paris: Five Qlients are connected through optical fibers to a Qonnector located in the SU campus. The length of the fiber links are 1~m for the link Alice-Qonnector, 3~km for the link Bob-Qonnector, 7~km for the link Charlie-Qonnector, 19~km for the link Dina-Qonnector and 31~km for the link Erika-Qonnector. Dutch Quantum City: Three Qlients connected through optical fibers to a Qonnector placed in Delft. The length of the fiber are 54~km for the link Fatou-Qonnector, 9~km for the link Geralt-Qonnector and 13~km for the link Hadi-Qonnector.}
    \label{fig:SatParisDelft}
\end{figure}

We consider the following network setting.
We suppose that two European Quantum Cities, the city of Paris with five Qlients and the `city' of the Netherlands with three Qlients are deployed, allowing for metropolitan scale quantum networking.  We also suppose that satellites are travelling over Europe, following an orbit allowing to send qubit states to the Qonnectors (see Fig.~\ref{fig:SatParisDelft}). The Qlients localisation corresponds to actual cities or laboratories. As in \cite{QCity}, the Paris Quantum City is composed of a Qonnector at the Sorbonne Université campus (SU), and 5 nodes: Sorbonne Université (SU-Alice),  Université Paris Cité campus (UPC-Bob), Orange Labs Châtillon (OR-Charlie), Télécom Paris (TP-Dina) and TGCC-CEA (CEA-Erika). The Dutch Quantum City is composed of a Qonnector at QuTech at the TU Delft campus and 3 Qlient nodes: Den Haag (Geralt), Rotterdam (Hadi) and Amsterdam (Fatou). Of course, please note that the choice of the Qonnector sites serves for illustration purposes only as these are urban areas, where installing optical ground stations with significant constraints would be challenging.\\

We recall below the baseline set of parameters used in our previous work~\cite{QCity} to simulate the performance of the Quantum City of Paris, which we will use here also to model sources, detectors and fiber links. These values are realistic with current or near term technology; see~\cite{QCity} and references therein for the justification of these parameters. \\

\begin{tabular}{l|c|l}
    $f_\text{qubit}$ & $\SI{80}{\mega\hertz}$ & Qubit creation attempt frequency\\
    $p_\text{qubit}$ & ${8\cdot10^{-3}}$  & Success probability of creation of a qubit  \\
    $p_\text{flip}$ & 0 & Flipping probability at the creation of a qubit \\
    $p_\text{crosstalk}$ & $10^{-5}$ & Probability that the detector flips the outcome \\
    $f_\text{EPR}$ & $\SI{80}{\mega\hertz}$ & EPR pair creation attempt frequency \\
    $p_\text{EPR}$ & $10^{-2}$ & Success probability of the creation of an EPR pair \\
    $p_\text{BSM}$ & 0.36 & Probability that a Bell state measurement succeeds \\
    $p_\text{transmit}$ & 0.81 & Probability that transmitting a qubit succeeds \\
    $t_\text{gate}$ & $\SI{1}{\nano\second}$& Time it takes to perform an operation on one qubit \\
    $p_\text{coupling}$ & 0.81 & Fiber coupling efficiency \\
    $\eta_\text{fiber}$ & $\SI{0.18}{\dB/\kilo\meter}$ & Fiber loss per kilometer \\ 
    $p_\text{dephase}$ & 0.02 & Phase flip probability in the fiber  \\ 
    $p_\text{det}$ &  0.95 & Detector efficiency (Probability that a measurement succeeds)  \\
    $R_\text{dark}$ & $10^2\si{\hertz}$ & Dark count rate \\
    $\Delta t_\text{det}$ & $\SI{100}{\pico\second}$ & Detector detection gate \\
    
\end{tabular}\\
\\\\
We point out that the rate at which single qubit states or EPR pairs are created depends strongly on the source model that we chose, namely Spontaneous-Parametric Down-Conversion (SPDC) in nonlinear crystals. This influences directly the rate at which entanglement can be created between different nodes of our network. This parameter, like all the others, can be tuned freely in our code for each source to match an actual source. For simplicity, we chose in this work to have the same qubit creation rates for the Qonnector and satellite sources. This is why we focus on the rate, in bit per attempt, at which protocols are performed instead of the throughput in bits per second. This gives a less source dependent view on the performance of the communication protocols that we study.\\

Using real live data from n2yo~\cite{n2yo}, a tracking website for satellites, and the orekit library~\cite{orekit}, a low level space dynamics library, we are able to find satellites in different orbits and to track down the precise time frame where they would pass over Europe. This also gives us other useful information such as the elevation of the satellite as well as the distance between the satellite and our ground stations at each point in time. In the following we will focus on four different satellite orbits: the QSS (Micius) orbit that was used in~\cite{MiciusSat}, the Starlink-1013 orbit, the Iridium-113 orbit and the Cosmos-2545 orbit and we focus on a time frame where the elevation of the satellites allows for quantum communication (set here at 20 degrees). The first two considered are low Earth orbit (LEO) satellites evolving at around 550~km above Earth, with slightly different orbits: the Micius orbit passes exactly above Paris while the Starlink orbit passes next to it. The Iridium satellite is higher, around 800~km above Earth. Last, the Cosmos satellite is a middle Earth orbit (MEO) satellite, at around 19000~km above Earth.\\

\begin{figure}[!ht]
 \begin{subfigure}{.5\textwidth}
   \centering
   \includegraphics[width=1.12\linewidth]{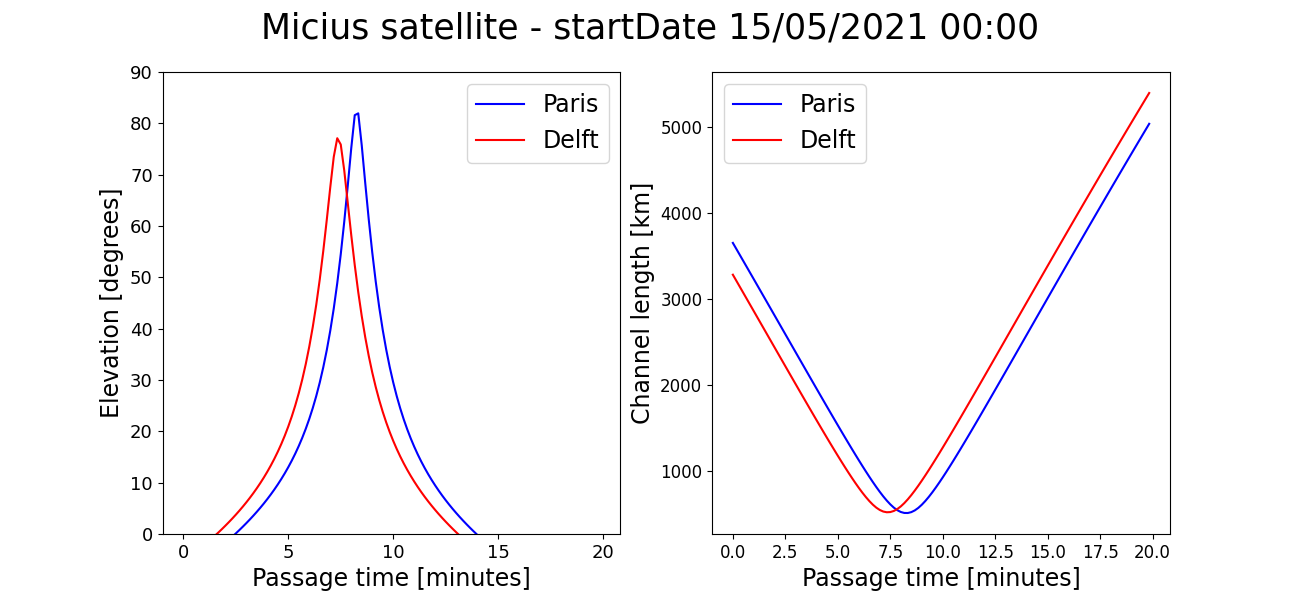}
    \caption{    }
   \label{fig:MiciusInfo}
 \end{subfigure}
  \begin{subfigure}{.5\textwidth}
   \centering
   \includegraphics[width=1.12\linewidth]{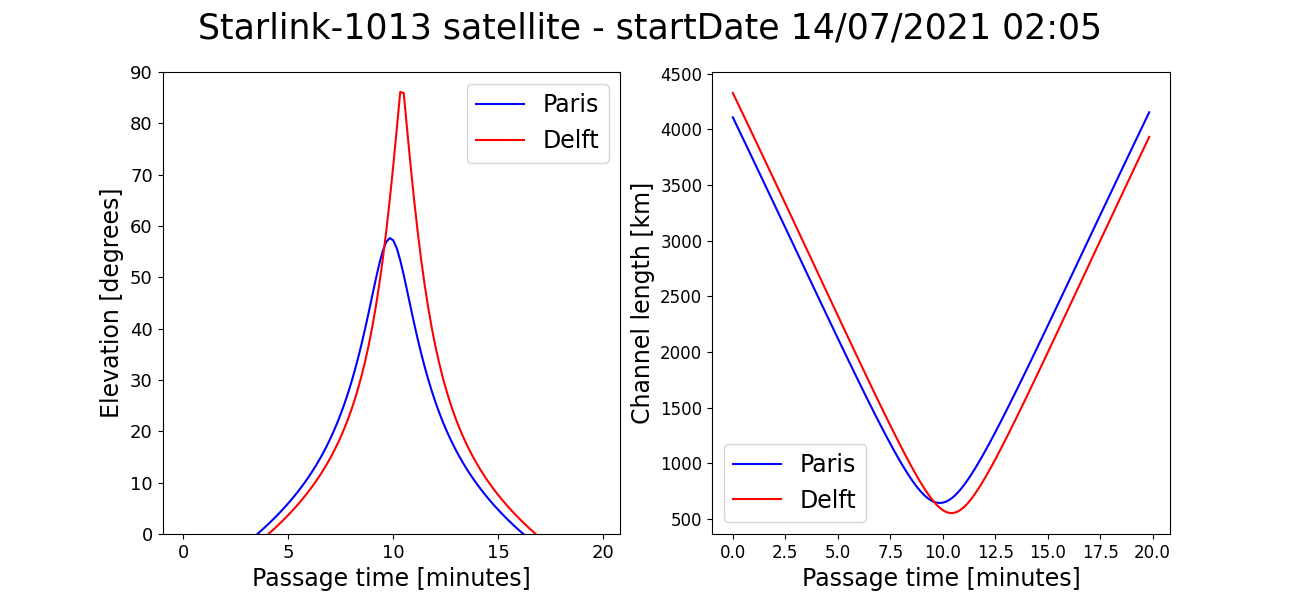}
   \caption{    }
   \label{fig:StarlinkInfo}
 \end{subfigure}
 
   \begin{subfigure}{.5\textwidth}
   \centering
   \includegraphics[width=1.12\linewidth]{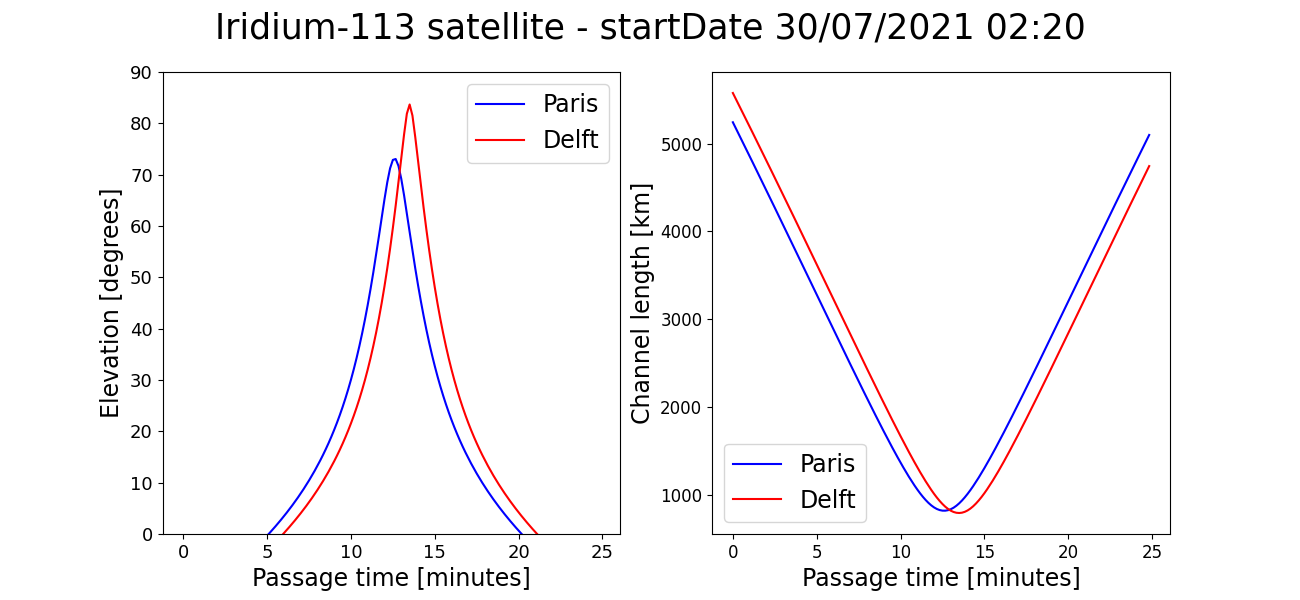}
   \caption{    }
   \label{fig:IridiumInfo}
 \end{subfigure}
   \begin{subfigure}{.5\textwidth}
   \centering
   \includegraphics[width=1.12\linewidth]{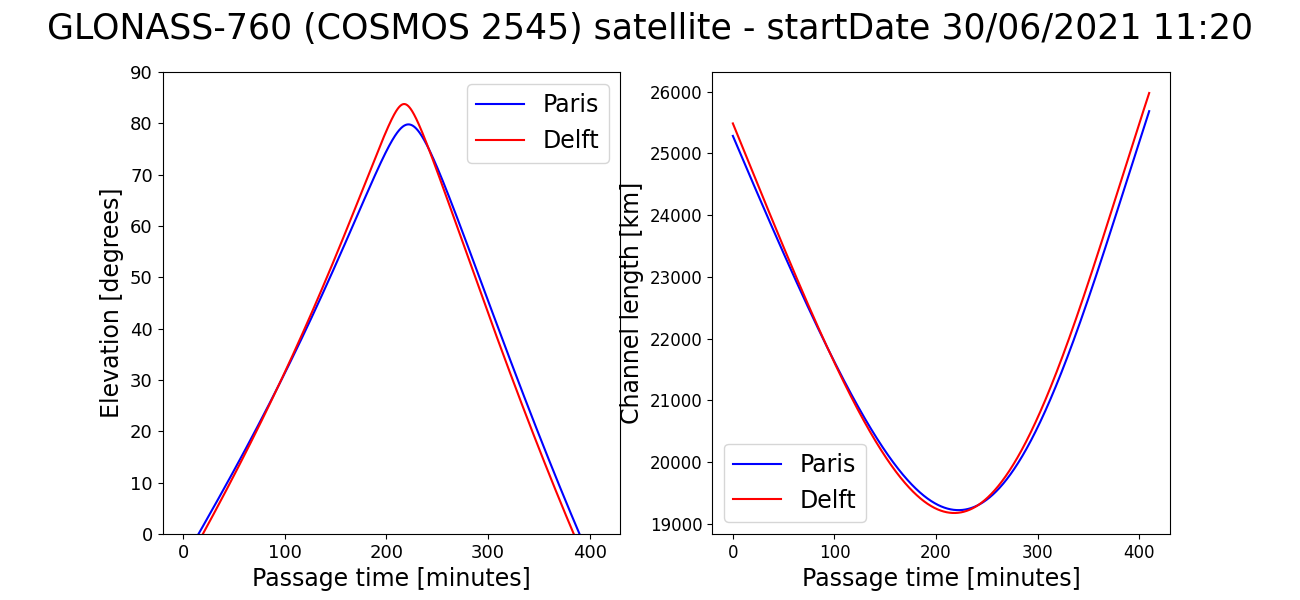}
   \caption{    }
   \label{fig:CosmosInfo}
 \end{subfigure}

 \caption{Elevation and distance to Paris and Delft of the (a) Micius,  (b) Starlink, (c) Iridium and (d) Cosmo satellites in the time frame considered.}
  \label{fig:SatInfo}
\end{figure}

In Fig.~\ref{fig:SatInfo} we show the elevation and the distance to the ground station for these satellites. We point out that our code is modular, accessible on GitHub~\cite{github,githubMatteo} and that any satellite can be investigated like we do in the following. 

\subsection{Simple downlink scenario: Choosing a satellite}
To test our model, we first simulate a simple downlink scenario between each of the satellites and the Qonnector from Paris. This will allow us to choose the satellite that is most suited for quantum communication.\\

\begin{figure}[!ht]
    \centering
    \includegraphics[width=0.9\textwidth]{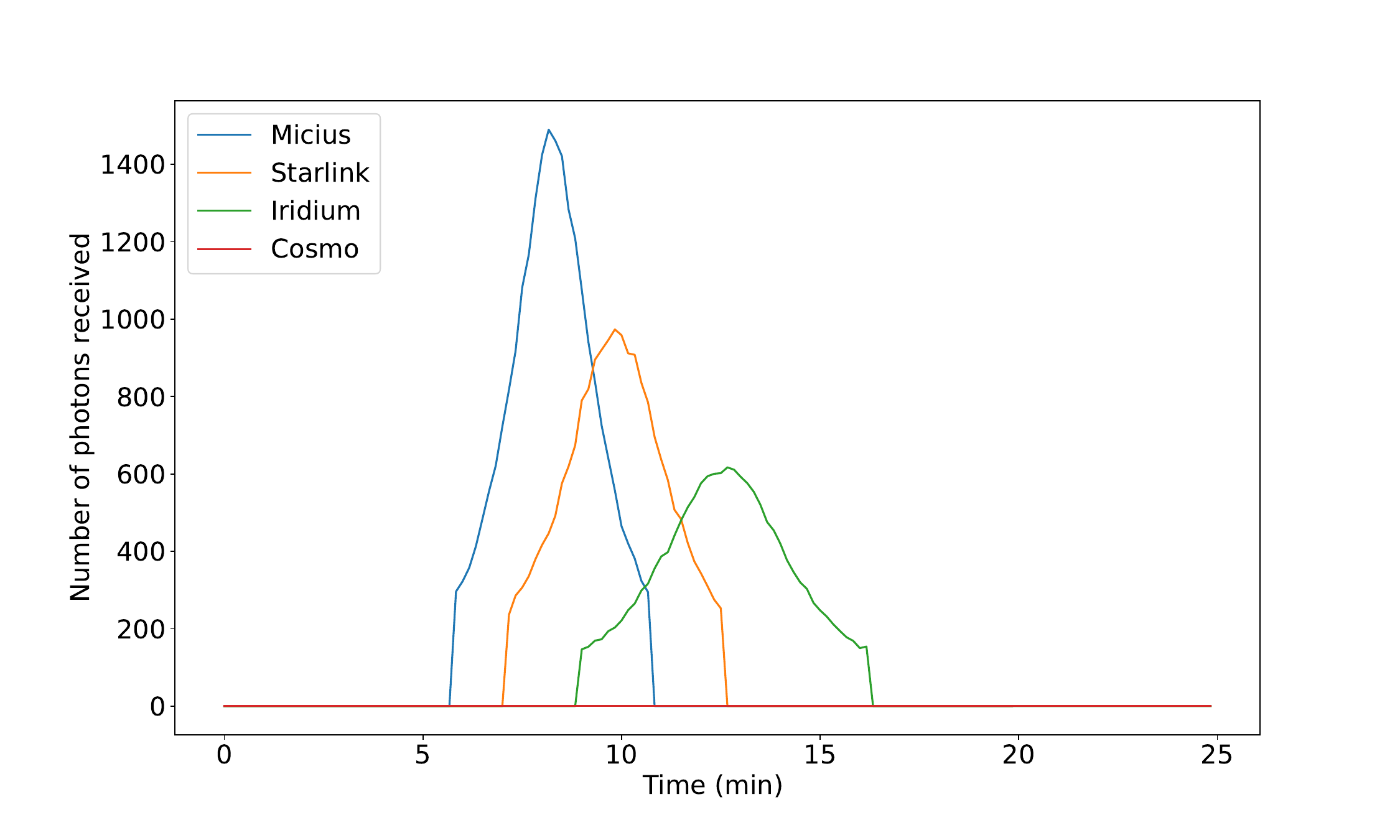}
    \caption{Comparison of the average number of photons received at the Paris Qonnector for the four satellites considered for approximately 6000 photons sent by the satellite at each point of its orbit. For this simulation we suppose there are no aerosols in the atmosphere and we set the aperture radius of the receiving telescope at 1~m, the beam waist divergence at $5$~$\mu$rad and the pointing error at $0.5$~$\mu$rad.
    }
    \label{fig:satcomp}
\end{figure}

The downlink scenario goes as follows: for each point in the orbit where the satellite's elevation is over 20 degrees, the satellite starts sending BB84 states to the Qonnector in Paris for one second while recording the time stamp of each state. The Qonnector receives, measures the states and records the measurement outputs. We can thus estimate the rate, i.e., the number of states received over the number of states sent, which also corresponds to the link efficiency at this point in the orbit. We average this over ten runs to get a better estimate. After this step is finished we move on to the next point in the orbit, ten seconds later. We show the result in Fig.~\ref{fig:satcomp} and in Table~\ref{table:SatComp} for a given set of parameters.\\

\begin{table}[!ht]
\centering
\begin{tabular}{c|c}

     Satellite & Maximum rate  \\
     \hline
     Micius & 0.238  \\
     Starlink & 0.157   \\
     Iridium & 0.101  \\
     Cosmo & 0 

\end{tabular}
     \caption{Maximum rate for the four satellites. The maximum rate is the rate, i.e., the number of qubits received at the ground station over the number of qubits sent from the satellite, when the satellite is the closest to the ground station.}
     \label{table:SatComp}
\end{table}

As expected, we observe that distance and elevation of the satellite impact the number of photons that arrive at the ground station. For example, we can see that none of the photons sent by the Cosmo satellite arrives at the Earth. The reason for this is that MEO satellites, while having the advantage of having a longer time frame during which the elevation allows for quantum communication than LEO satellites, are too far for single-photon states to arrive at a precise point on Earth due to the combined effects of pointing errors and beam wandering. It follows that this is even more challenging for geostationary satellites, which are at a height of ~36000~km above ground and have the significant advantage of always visible by a given ground station. As it is located further than the other two satellites, the Iridium satellite has a lower rate, but a longer exploitation time. We thus identify the well-known trade-off in satellite communication between distance of the satellite to the Earth and time frame in which we can use it, which we will discuss in more detail in Sec.~\ref{sec:discussion}. The Micius and Starlink satellites are performing better but as can be seen in Fig.~\ref{fig:StarlinkInfo}, the elevation angle of Starlink with respect to the Paris ground station is lower than the one of Micius. This means Starlink does not pass exactly above Paris, which causes a drop in the rate.\\   


\subsection{Influence of the parameters}
As detailed in Sec.~\ref{subsec:SatModel}, our loss model allows us to analyse the effect of a few important parameters of satellite communication, namely the aperture radius of the receiving telescope, the beam waist divergence, the pointing error and the aerosol model, which affect directly the atmospheric transmittance. In this section we study the effect of these different parameters on the rate in the downlink scenario. Based on the previous analysis, we chose to focus on the Micius satellite for the rest of this section as it exhibits the highest rate with the Paris node. When studying a parameter, we fix the other ones to what we expect to be the best value achievable in the near future: an aperture radius of the receiving telescope of 1~m, a beam waist divergence at $5$~$\mu$rad and a pointing error at $0.5$~$\mu$rad. We also suppose that the turbulence induced by beam-wandering is negligible with respect to the pointing error~\cite{LuisVictorFeasability}.\\

We start with a study of the atmospheric model. In Sec.~\ref{subsec:SatModel} we have detailed a few atmospheric models that have an effect on photonic communication in free-space, and that have different meteorological ranges. The meteorological range is usually defined as the length of atmosphere over which a beam of light travels before its luminous flux is reduced to 2\% of its original value~\cite{Kneizys1988}. Here we will study the ideal case where there is no aerosol between the ground station and the satellite, the \textsf{rural5} and \textsf{rural23} models that correspond to ground stations in rural areas with a meteorological range of, respectively, 5 and 23 km, the \textsf{urban5} model that correspond to a ground station close to a city and the \textsf{navy} model corresponding to a ground station in the middle of the sea. \\

\begin{figure}[!ht]
    \centering
    \includegraphics[width=13cm]{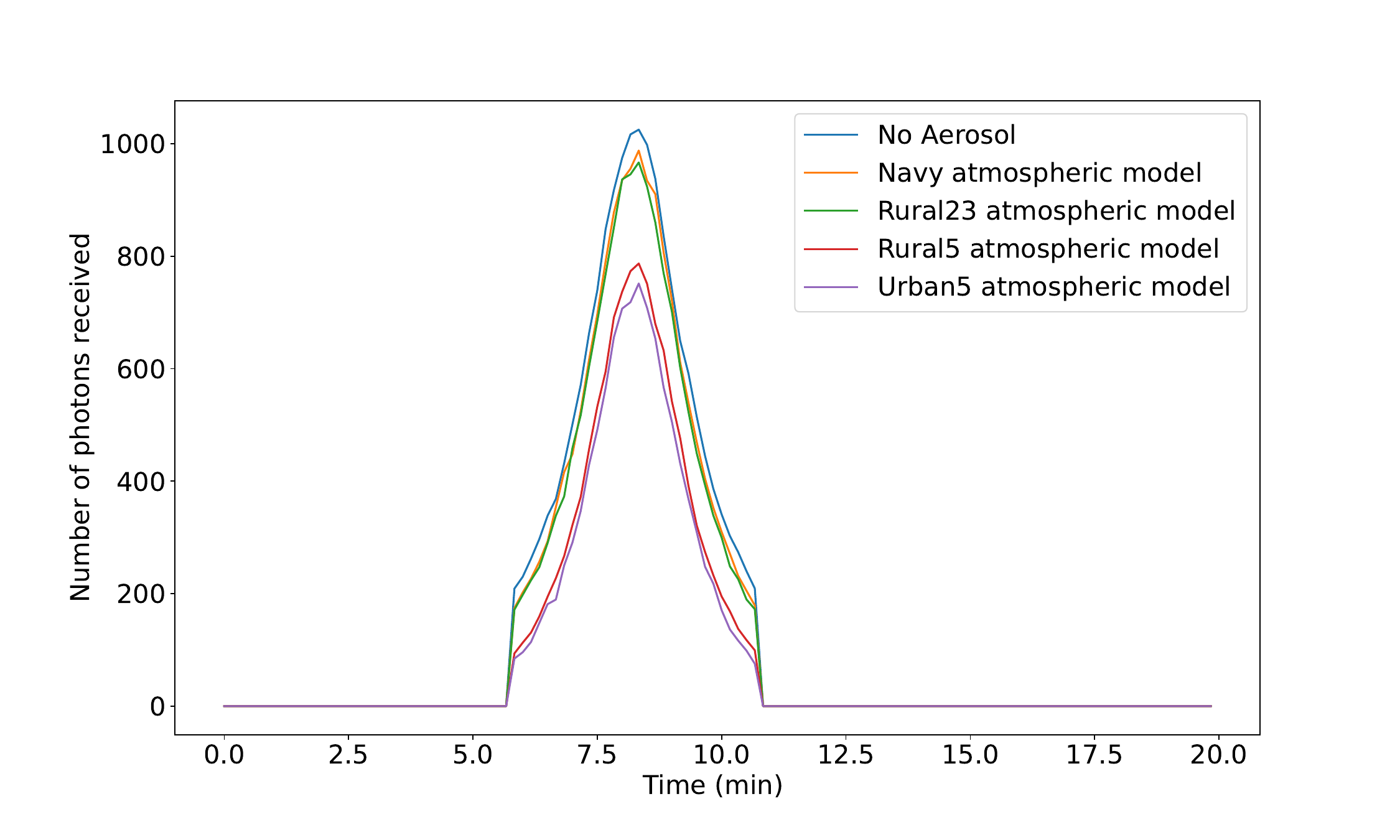}
    \caption{Effect of different aerosol models on the number of photonic states arriving from the Micius satellite.}
    \label{fig:AtmComp}
\end{figure}

In Fig.~\ref{fig:AtmComp} we show the number of photons received at the ground station when considering these different atmospheric models. This simulation has been done similarly to the one in the previous section, meaning that BB84 states are sent by the satellite for each point of the orbit where the elevation of the satellite is over 20 degrees. As expected, we observe in the figure that the smaller the meteorological range, the more photons are lost in the atmosphere. However, as we will see in Sec.~\ref{sec:discussion}, atmospheric parameters have more drastic effects on the photon transmission when we get closer to the ground as aerosols are mostly concentrated there.\\

\begin{figure}[!ht]\hspace*{-3cm}
 \begin{subfigure}{.425\textwidth}
   \centering
   \includegraphics[width=1.1\linewidth]{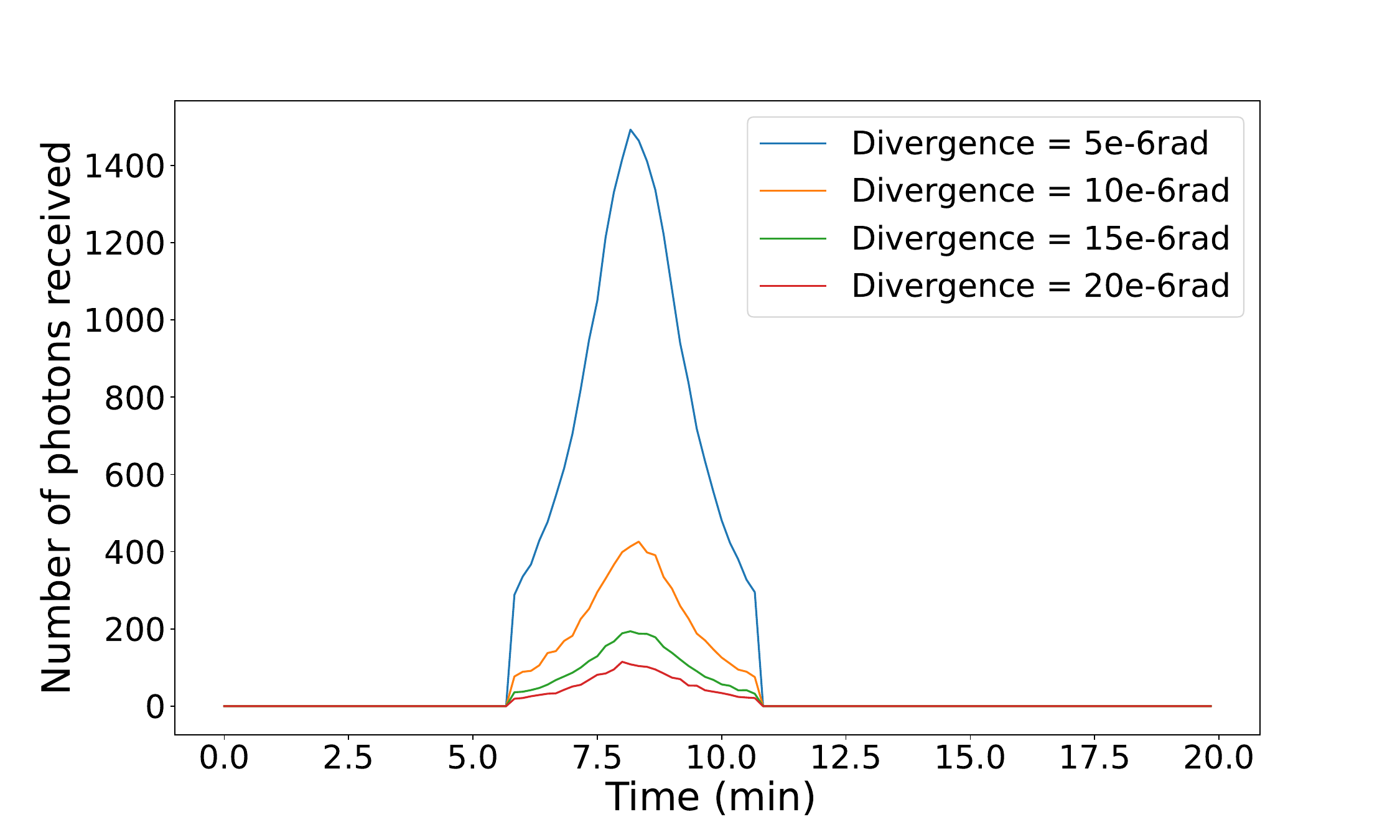}
   \caption{}
   \label{fig:DivComp}
 \end{subfigure}
    \begin{subfigure}{.425\textwidth}
   \centering
   \includegraphics[width=1.1\linewidth]{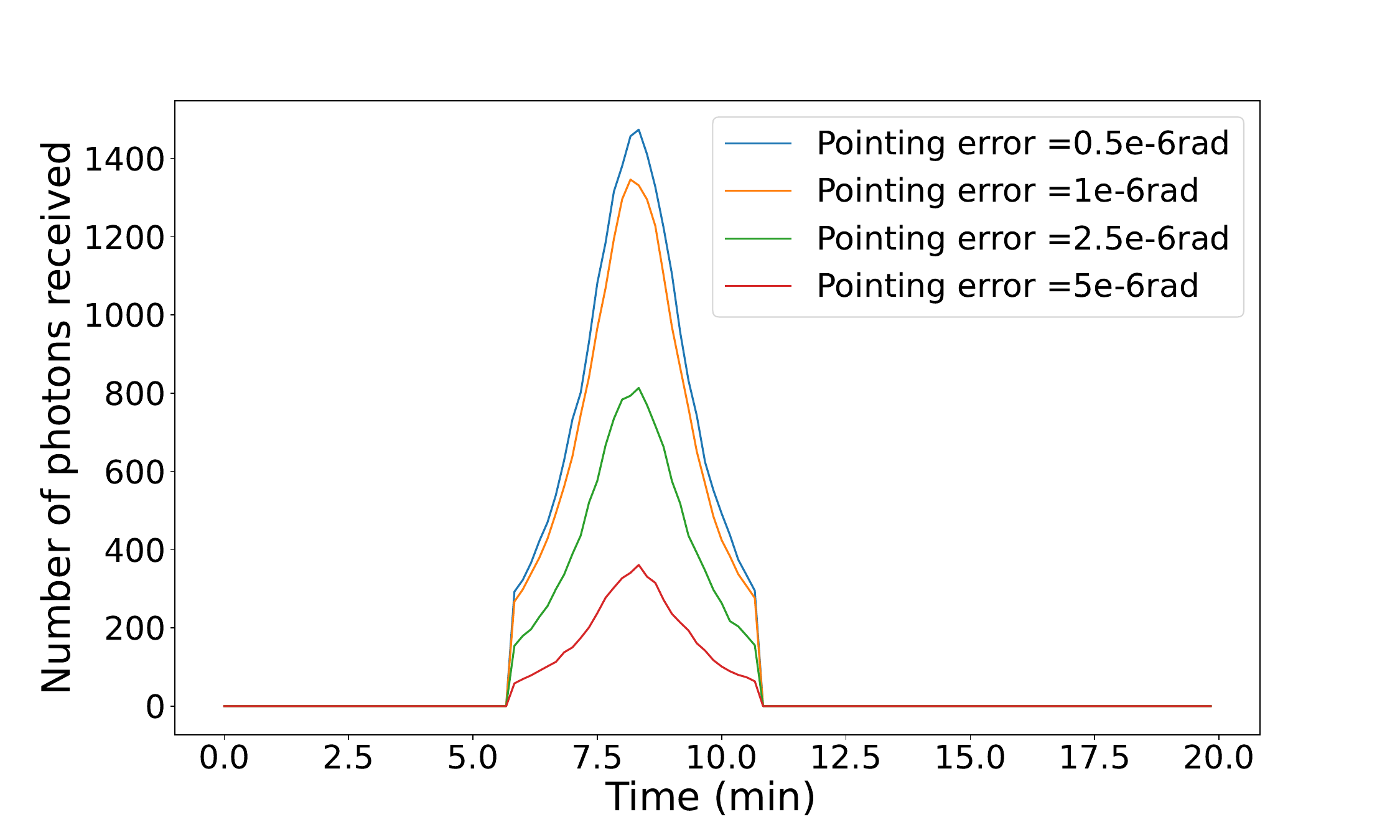}
   \caption{}
   \label{fig:PointComp}
 \end{subfigure}
  \begin{subfigure}{.425\textwidth}
   \centering
   \includegraphics[width=1.1\linewidth]{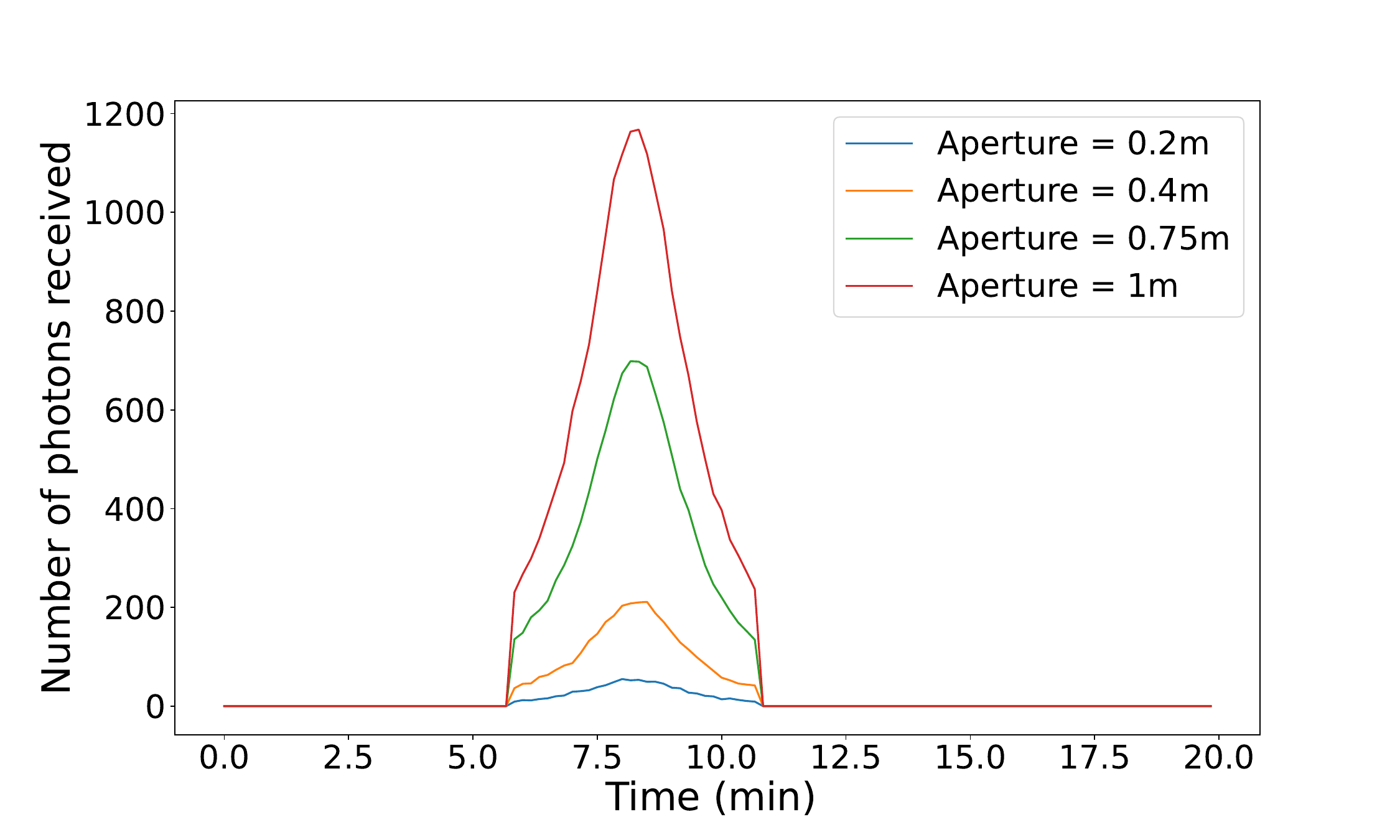}
   \caption{}
   \label{fig:RxComp}
 \end{subfigure}

\caption{Number of photons received, from Micius satellite to ground station.
In each plot, we fix non studied values as baseline values: no atmospheric turbulence, an aperture radius of the receiving telescope of 1~m, a beam divergence at $5$~$\mu$rad and a pointing error at $0.5$~$\mu$rad.
The three subfigures show the effect of individually varying one of these parameters, while keeping the others constant: (a) the beam divergence angle $\theta_d$, (b) the pointing error $\theta_p$   and (c) the aperture radius of the receiving telescope.} 
\label{fig:ParamComp}
\end{figure}

We then study, using again the downlink scenario, the effect of modifying the other parameters included in our model. This study allows us to know what to expect from a specific setting and to identify the key parameters to improve future quantum communication. The results are shown in Fig.~\ref{fig:ParamComp}.
We see that the transmitter parameters do not all have the same importance. A change of $5$~$\mu$rad in the beam divergence angle (see Fig.~\ref{fig:DivComp}), has a drastic effect on the number of photons arriving at the ground station. On the other hand, the pointing error of the transmitter (see Fig.~\ref{fig:PointComp}) has to be multiplied by five in order to reduce the number of arriving photons by a half. Finally, we see that increasing the aperture radius of the receiving telescope at the ground station by $20$~cm can almost double the number of qubits successfully measured (see Fig.~\ref{fig:RxComp}). This last parameter might be the easiest to improve in future experimental realizations.

\section{Quantum Key Distribution between two Qlients}
\label{sec:QKD}

We now embed this study in the context of quantum networks by showing the performance of satellite communication when linking two Quantum Cities. We study the achievable QKD rate between two Qlients respectively from the quantum city of Paris and the Dutch quantum city connected via the Micius satellite (see Fig.~\ref{fig:SatParisDelft} and \ref{fig:MiciusInfo}). We will consider the baseline set of parameters from the previous section, namely no atmospheric turbulence, an aperture radius of the receiving telescope of 1~m, a beam divergence at $5$~$\mu$rad and a pointing error at $0.5$~$\mu$rad.\\

Let us imagine that one Qlient from the Paris Quantum City, say Bob, wants to share a secret key with a Qlient from the Dutch Quantum City, say Hadi, using a QKD protocol. There are different ways to achieve this functionality and we focus our analysis on two of them. For a more extensive study of the different ways of achieving QKD as well as some other protocols in a metropolitan network, see~\cite{QCity}. In the following we show the feasibility of two QKD scenarios, with a trusted and an untrusted node, and we then discuss their practical relevance.

\subsection{Trusted satellite}
\label{sec:trustednode}

The most standard way to achieve quantum key distribution between Bob and Hadi via the satellite is to perform BB84 between these nodes while trusting each of them not to reveal the keys they generate. More precisely the satellite, when above one of the ground stations, performs two BB84 protocols in parallel to establish two keys with the Qonnectors of the two cities, $k_{\textrm{Paris}}$ and $k_{\textrm{Delft}}$. At the same time the two Qlients establish secret keys, $k_{\textrm{Bob}}$ and $k_{\textrm{Hadi}}$ using the BB84 protocol with their Qonnector. Once all keys are created, the Paris Qonnector can send $k_{\textrm{Bob}}$ as message to the satellite using $k_{\textrm{Paris}}$ as key (with a classical encryption scheme such as one-time pad), and the satellite forwards the message to the Delft Qonnector using the key $k_{\textrm{Delft}}$, which subsequently sends the message to Hadi using the key $k_{\textrm{Hadi}}$. In the end, Bob and Hadi share $k_{\textrm{Bob}}$.\\

\begin{figure}[!ht]
    \centering
    \includegraphics[width=16cm]{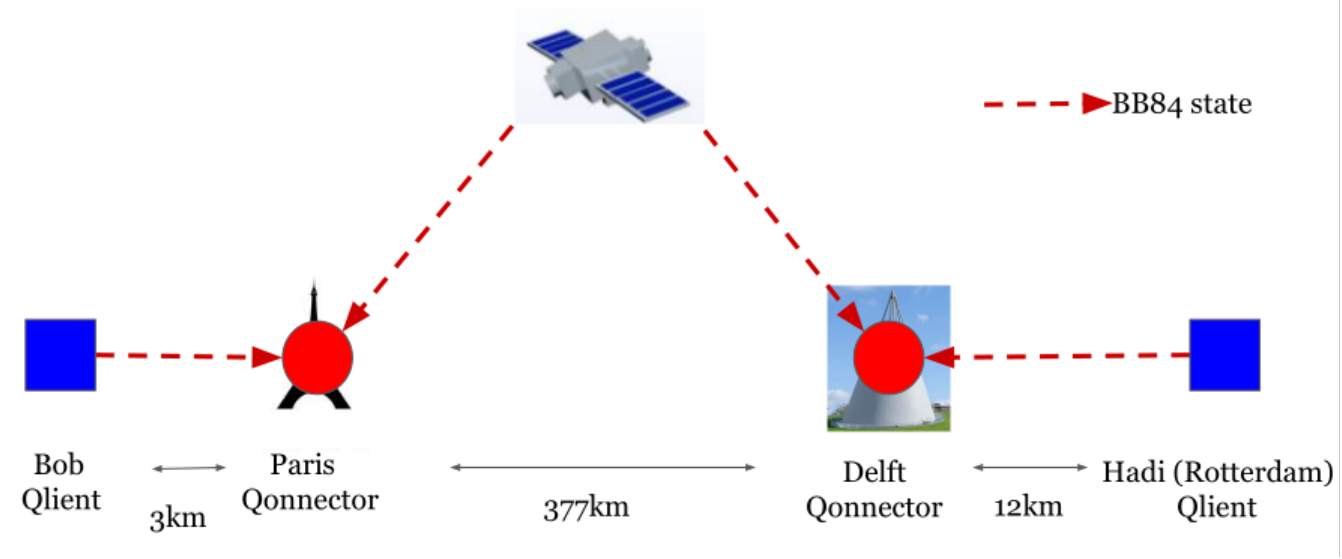}
    \caption{QKD between Bob and Hadi: All trusted node scenario. Here keys are generated between all different node pairs and are transfered as secret messages. 
    In the end, Bob and Hadi share $k_{\textrm{Bob}}.$}
    \label{fig:ScenarAllBB84}
\end{figure}

\begin{figure}[!ht]
    \centering
    \includegraphics[width=14cm]{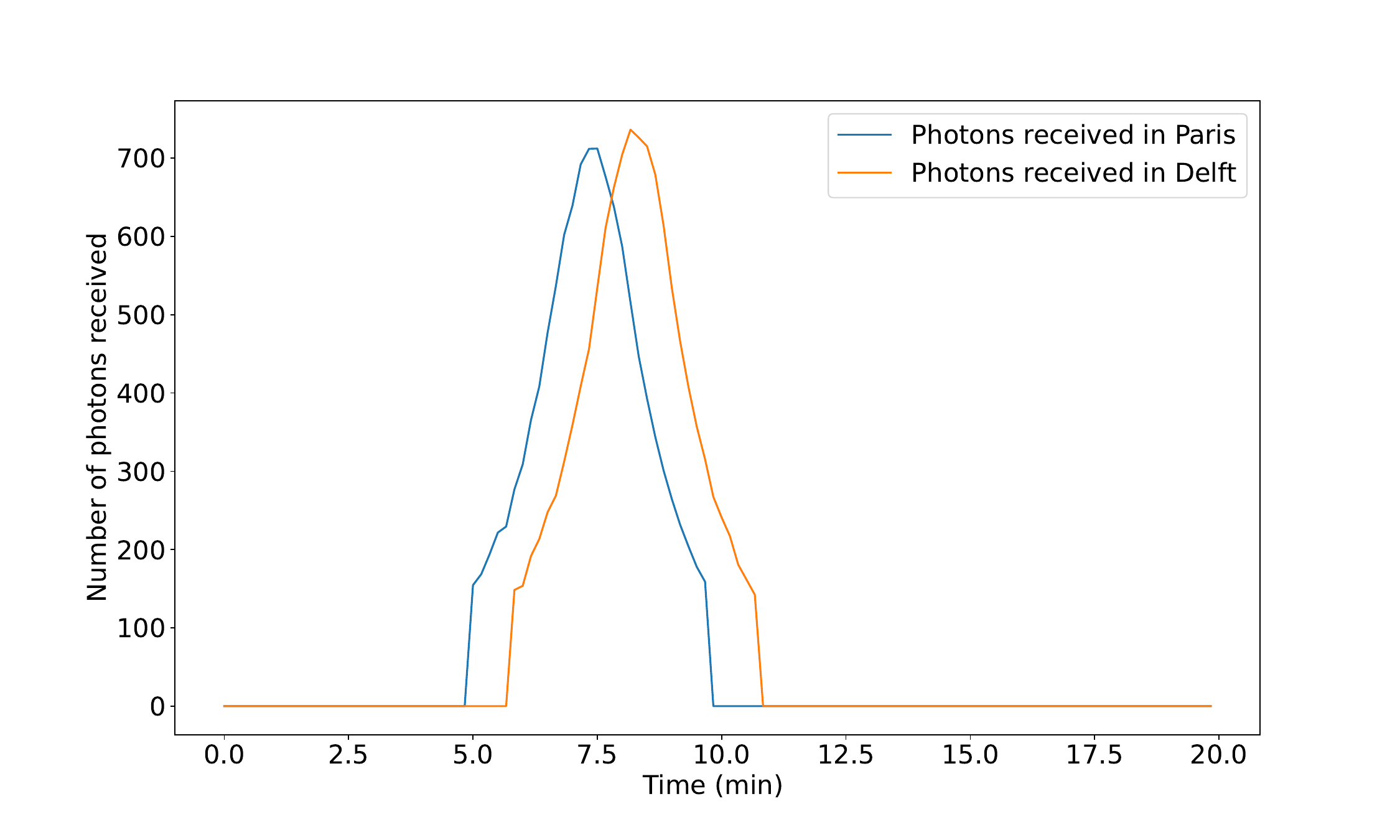}
    \caption{Average number of photons received in Paris and Delft for approximately 3000 photons sent from the satellite at each orbit point.}
    \label{fig:downlinkboth}
\end{figure}

As in the analysis in the previous section, we simulate sending BB84 states to both ground stations and obtain the rate for the two satellite to ground links. We also simulate sending BB84 states from each Qlient to the Qonnectors. The parameters for all nodes are listed in Sec.~\ref{sec:param}. We show the results of these simulations in Fig.~\ref{fig:downlinkboth} and Table~\ref{tab:BB84forth}.\\

\begin{table}[!ht]
    \centering
    \begin{tabular}{|c|c|}
    \hline
         Nodes involved  & Rate (raw key bit per channel use)\\ \hline
         Satellite $->$Paris Qonn & 0.238 \\
         Satellite $->$ Dutch Qonn & 0.228\\ \hline
         Alice $->$ Paris Qonn & 0.423  \\
         Bob $->$ Paris Qonn &  0.374  \\
         Charlie $->$ Paris Qonn & 0.322  \\
         Dina $->$ Paris Qonn &  0.180  \\
         Erika $->$ Paris Qonn &  0.115  \\ \hline
         Fatou $->$ Dutch Qonn & 0.043\\
         Geralt $->$ Dutch Qonn & 0.296\\
         Hadi $->$ Dutch Qonn & 0.253\\ \hline

    \end{tabular}
    \caption{Performance of the BB84 protocol between all nodes. The first two lines correspond to the satellite sending BB84 states to the two ground stations, and the other lines correspond to BB84 rates with each Qlient inside the two Quantum Cities. }
    \label{tab:BB84forth}
\end{table}

We observe, as expected, that the longer a photon has to travel in a fiber the lower is the rate. These number were computed with the satellite exactly above the Paris ground station, which explains the small difference in rate in the first two lines. In this scenario, the rate of the overall key distribution protocol is given by the minimum rate over all sublinks. Hence depending on the pair of nodes that want to establish a secure key, the limiting sublink can be either the satellite-to-ground link or the fiber link. For example if Hadi and Bob want to perform the QKD protocol considered here, their total rate is limited by the rate of the satellite-to-ground link. But if Erika and Fatou want to do the same, it is the fiber link between Fatou and her Qonnector that is most limiting. Note that the satellite-to-ground rate here is the rate when the satellite is at its optimal position, namely just above the ground stations. As an example of performance, with a source rate of \SI{80}{\mega\hertz}, the raw key rate at this point of the orbit is 1.7~Mbit/s for Bob and Hadi and  300~kbit/s for Erika and Fatou. We discuss how practical these values are in Sec.~\ref{subsec:RealQKD}. Note once more that in order for this scenario to securely create a key between two Qlients, all the nodes (Qonnectors and satellite) along the path between two Qlients have to be trusted.

\subsection{Untrusted satellite}
\begin{figure}[!ht]
    \centering
    \includegraphics[width=16cm]{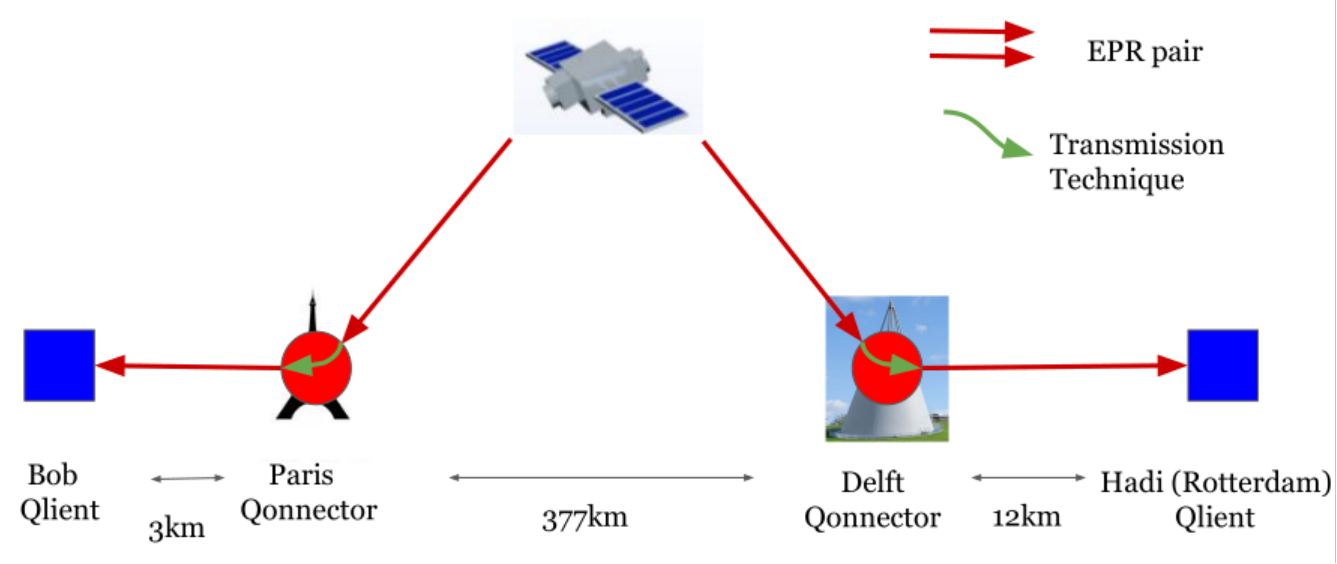}
    \caption{QKD between Bob and Hadi: Untrusted nodes. Here the satellite performs the BBM92 protocol directly with the Qlients. The Qonnectors are used as transmitting stations that couple the photons arriving from the satellite into a fiber and send them to the Qlients.}
    \label{fig:ScenarEPRTrans}
\end{figure}

Another way to distribute a secret key between Hadi and Bob is to use an entanglement-based version of QKD, and in particular the BBM92 protocol~\cite{BBM92}. This requires that an EPR pair is shared between the two Qlients. More specifically, in the context of Quantum Cities connected by a satellite, the protocol goes as follows (see Fig.~\ref{fig:ScenarEPRTrans}): an EPR pair in the state $\ket{\psi^-}=\frac{1}{\sqrt{2}}(\ket{01}-\ket{10})$ is created at the satellite and each qubit of the pair is sent towards two Qonnectors at the ground. It is then coupled into an optical fiber and transmitted to the Qlient who measures it. The Qlients keep the outcomes measured with the same timestamp to post-select the qubits that came from the same pair. These steps are performed for each point of the Micius orbit where the satellite elevation is above 20~degrees for both Qonnectors. The coupling of a photon coming from a satellite into a fiber succeeds with probability $p_\text{transmit}$ that we fix to 81\% ~\cite{couplingfiber}. Coupling photons arriving from a satellite into a fiber can be improved by using adaptive optics to correct atmospheric effects~\cite{Valentinarticle}. This parameter is of course freely tunable in our simulation.\\


\begin{figure}[!ht]
    \centering
    \includegraphics[width=11cm]{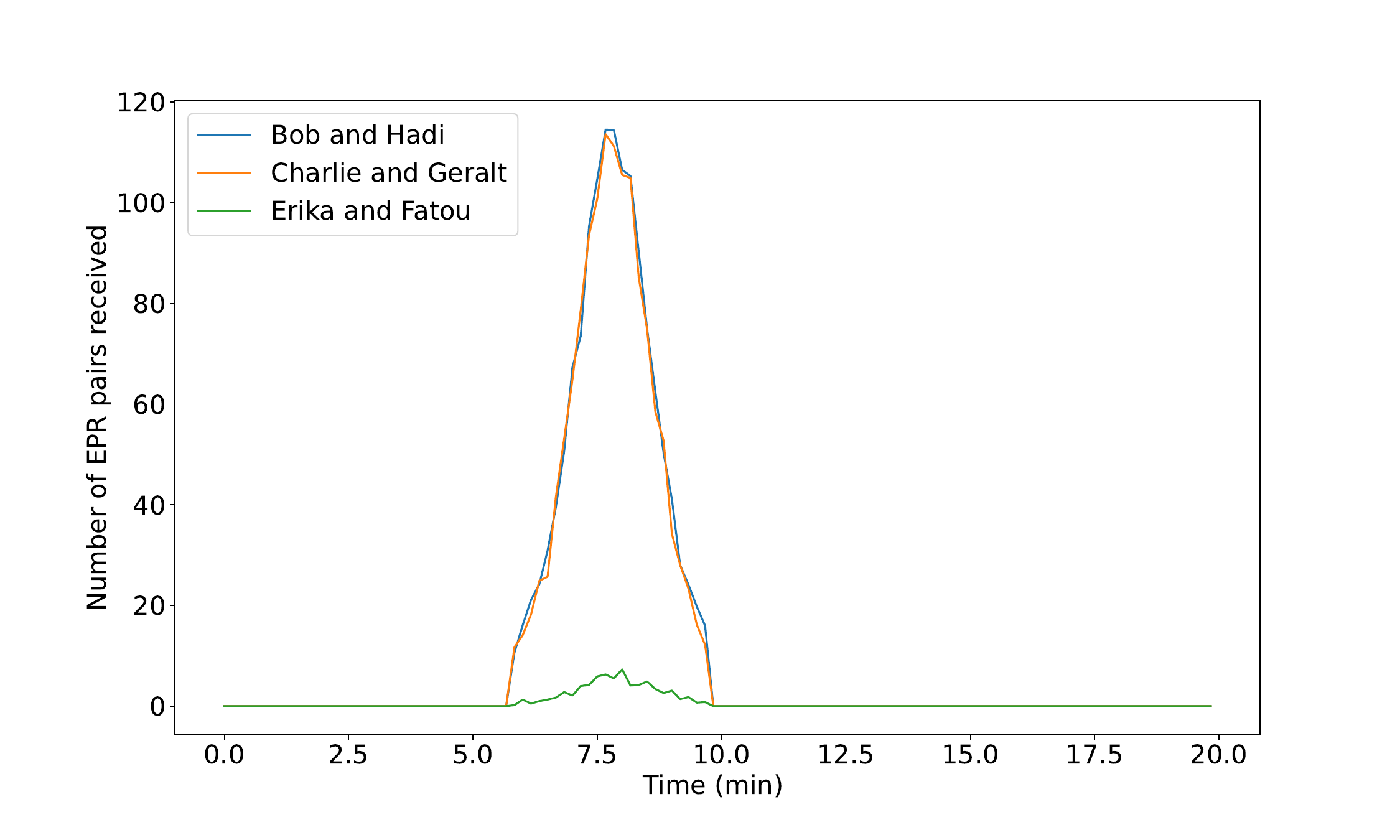}
    \caption{Average number of successfully transmitted EPR pairs from the Micius satellite to pairs of Qlients in the two Quantum Cities considered. For each point of the orbit we simulate sending approximately 650 EPR pairs from the satellite and average over 10 rounds.}
    \label{fig:EPRtrans}
\end{figure}

Simulating the process described above for each point in the satellite orbit and averaging over tens of runs, we obtain the rate as the ratio between the number of pairs sent from the satellite and the number of pairs received by the Qlients. We show the results of simulations with different pairs of Qlients in Fig.~\ref{fig:EPRtrans} and Table~\ref{table:EPRtrans}. 

\begin{table}[!ht]
\centering
\begin{tabular}{c|c}

     Qlient pairs & Rate (raw key bit per channel use) \\
     \hline
     Bob \& Hadi & 0.0183  \\
     Charlie \& Geralt & 0.0185  \\
     Erika \& Fatou & 0.0019 \\

\end{tabular}
     \caption{Average rate for pairs of Qlients of the two Quantum Cities considered when the satellite is exactly in the middle of the two ground stations.}
     \label{table:EPRtrans}
\end{table}

As before, in addition to the loss due to atmospheric perturbation between the satellite and the ground stations, the rate is affected by the distance photons have to travel in optical fibres from the Qonnectors to each Qlient. As expected the longer this distance is, e.g., for Erika and Fatou, the lower the rate is. This is also why the rate is similar for the two pairs of Qlients Bob \& Hadi and Charlie \& Geralt: the distance photons have to travel into fibers is comparable (see Fig.~\ref{fig:SatParisDelft}). With a source rate of \SI{80}{\mega\hertz}, the raw key throughput would be 150~kbit/s for Hadi and Bob, and 14~kbit/s for Erika and Fatou.\\

We emphasize again that compared to this untrusted node scenario, the trusted node scenario as studied in the previous section has a higher rate, albeit at the cost of the additional security assumption that the Qonnector and satellite nodes are not under control of an adversary.

\subsection{Realistic quantum key distribution}
\label{subsec:RealQKD}
The above simulations illustrate the feasibility of quantum key distribution between two Quantum Cities separated by a few hundreds of kilometers. However this analysis assumes that the satellite is always visible by the ground stations. This is unrealistic since, since as we saw in Sec.~\ref{subsec:SatModel}, the time span during which a satellite's elevation allows for quantum communication is only a few minutes. Considering this, in the trusted node scenario above, the total raw key established between the Micius satellite and the Paris ground station during the whole passage of the satellite above Europe is of length 17~kbits on average. This limits significantly the amount of messages that can be securely sent between the two Qlients considered.\\

To overcome this limitation, a possible solution is to use multiple passages of a satellite over several days to create and store keys at the ground stations. These keys can be established, for instance, using one of the two scenarios presented above. They can subsequently be used to establish a secure communication channel between the Qlients. This is useful in scenarios where relatively small amount of shared secret key is required. The secure storage of the key material at the nodes also needs to be considered in this case.
 
 This solution may not however apply to quantum communication protocols beyond QKD, where nodes need to be remained entangled for long periods of time. 
Constellations of satellites orbiting around the Earth offer an alternative solution. In this case, once a satellite is out of reach for quantum communication, another suitable satellite in the constellation can be used to continue the key generation. Finding the optimal orbit height and the optimal number of satellites on this orbit is a complex question and detailed studies are required to address all the associated challenges; see for example~\cite{SatConstellation}.\\

Some other issues that may directly affect the key rate have not been taken into account in this study. For example, we neglected here the variations of atmospheric turbulence with the position of the Qonnector and the conditions at the time of the key establishment, which play a central role in satellite communication. Moreover, in our simulations we considered the wavelength of the photon to be 1550~nm, which is convenient for coupling them with telecom equipment on the ground. In the Micius experiment~\cite{MiciusSat}, the operational wavelength is 850~nm, which would also alter the results. Finally, as we focus in study on the feasibility of satellite communication in the context of quantum networks, we have neglected entirely the problem of scheduling operations at each node. In realistic scenarios, the Qonnector and satellite nodes may not be able to perform operations in parallel and thus would have to store qubits until their processing unit is available. Including the effect of quantum memories and synchronization techniques is a crucial follow-up work of the present analysis.\\

Even without covering all aspects of a full quantum communication network,  we hope that this study shows the possibilities and limitations of satellite communication to link local networks. According to our simulations, current technologies could already allow for interesting applications between distant cities, as we discuss in the next section.


\section{Discussion}
\label{sec:discussion}
\subsection{Comparison with ground-based and balloon-based communication}

As we mentioned in the introduction, current quantum repeater technologies do not allow for practical key rates over the distances that we consider here. Assuming the realization of a quantum repeater link over about 50~km in the next few years, since the distance between the Dutch Quantum City and the one of Paris is 377~km, we would need between five and ten repeater nodes. Important quantum repeater parameters include the overall efficiency of the link (comprising the probability of photon emission, the storage-and-retrieval efficiency, the coupling and detection efficiencies), and the probability of success of the swapping operation via the Bell state measurement. Promising technologies based on various platforms~\cite{PompiliScience2021,Multiplex2,Lago_Rivera_2021} or alternative models such as all photonic quantum repeaters~\cite{AllPhotonic} are presently under intense investigation to achieve this goal.\\

\begin{figure}[!ht]
    \centering
    \includegraphics[width = 13cm]{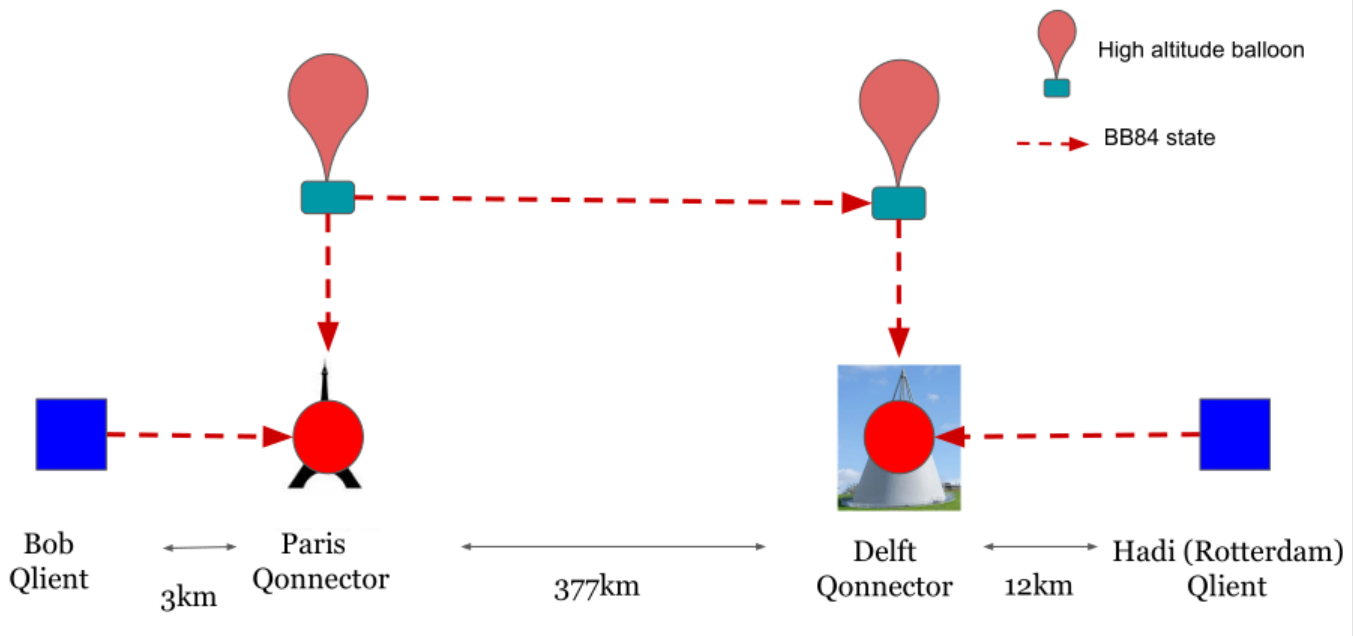}
    \caption{A trusted node QKD scenario between two Qlients using two high-altitude balloons over the Qonnectors. We assume that the balloons are located directly above each Qonnector.}
    \label{fig:drone}
\end{figure}

In parallel, an alternative technological path that could be envisioned is the use of drones or high-altitude balloons. Let us imagine a quantum key distribution scenario using two stationary high-altitude balloons above the Paris and Dutch Qonnectors. As we show in Fig.~\ref{fig:drone}, this configuration is the same as the one considered in Sec.~\ref{sec:trustednode}, whereby secret keys are established between each pair of nodes and used to transmit as secret message the final private key between the two Qlients.
In the rest of this section, we explore some key parameters in order to obtain an estimate of the key rate in this scenario.

Using the free-space loss model described in Sec.~\ref{subsec:SatModel} and the parameters of Sec.~\ref{sec:param}, we obtain the key rate between the two balloons and between each balloon and the Qonnector using our simulation tools. First, the height of the balloons is of crucial importance because the atmospheric transmittance gets smaller when closer to the ground as we show in Table~\ref{tab:Tatmstudy}. Moreover, the aerosols are concentrated in the lower layers of the atmosphere and thus have a much higher impact on the atmospheric transmittance when the photon path is closer to the ground as we show in Table~\ref{tab:Aerostudy}. We recall that the \textsf{rural5} and \textsf{rural23} aerosol models correspond to ground stations in rural areas with a meteorological range of, respectively, 5 and 23 km, and the \textsf{urban5} aerosol model corresponds to a ground station close to a city.\\

\begin{table}[!ht]
    \centering
    \begin{tabular}{c|c}
    Height  & $T_\text{atm}$ \\ \hline
         10~km & 0.96753  \\
        5~km & 0.85255 \\
        1~km & 0.26363\\
    \end{tabular}
    \caption{Atmospheric transmittance for different heights above the ground.}
    \label{tab:Tatmstudy}
\end{table}

\begin{table}[!ht]
    \centering
    \begin{tabular}{c|c|c}
    Aerosol model & $T_\text{atm}$ at 10~km & $T_\text{atm}$ at 1~km  \\ \hline
    No aerosol & 0.96753 & 0.26363 \\
    \textsf{rural23} model &  0.90658 & 1.6209e-7 \\
    \textsf{rural5} model &  0.90647 & 1.4159e-31 \\
    \textsf{urban5} model &  0.906622 & 3.2276e-38 \\
    \end{tabular}
    \caption{Atmospheric transmittance at an altitude of 10~km and 1~km for different aerosol models.}
    \label{tab:Aerostudy}
\end{table}

For our study here, we fix the balloons height at 10~km and suppose that there are no aerosols in the atmosphere. We also suppose that the aperture radius of the receiving telescope in the balloons is 40~cm, and we fix the beam divergence at $5$~$\mu$rad and the pointing error at $0.5$~$\mu$rad. We compute separately the horizontal atmospheric transmittance between the two balloons and the vertical atmospheric transmittance between the balloon and the ground station (around 0.9). As the free-space communication happens in the atmosphere, we can no longer neglect the effect of the refractive index structure constant $C_n^2$ as it was done in the satellite case. In reality, the value of $C_n^2$ also depends on the height of the high-altitude balloons. As we show in Table~\ref{tab:Cn2study}, this value has a drastic effect on the key rate between the two balloons. In what follows we choose an optimistic yet non-zero value for $C_n^2$ between the two balloons, namely $10^{-17}$. For the balloon-to-Qonnector links we fix $C_n^2$ at $10^{-15}$ and take the aperture of the receiving telescope at the Qonnector to be 1~m. In Table~\ref{tab:dronebobtohadi} we show the rate for all the sublinks between Bob and Hadi.\\

\begin{table}[!ht]
    \centering
    \begin{tabular}{c|c}
    Value of $C_n^2$ & Rate \\ \hline
         0 & 0.138  \\
        $10^{-17}$ & 0.079 \\
        $10^{-16}$  & 0.014\\
        $10^{-15}$ & 0.001 \\
        $10^{-13}$ & 1e-5 \\
    \end{tabular}
    \caption{BB84 rate for balloon-to-balloon communication for different values of $C_n^2$. The balloons are separated by 377~km and are located at a height of 10~km.}
    \label{tab:Cn2study}
\end{table}

\begin{table}[!ht]
    \centering
    \begin{tabular}{c|c}
    Sublink & Rate \\ \hline
         Bob $->$ Paris Qonn & 0.374   \\
         Paris Drone $->$ Paris Qonn & 0.463\\
        Paris Drone $->$ Dutch Drone & 0.079 \\
        Dutch Drone $->$ Dutch Qonn & 0.463 \\
        Dutch Qonn $->$ Hadi & 0.253\\
    \end{tabular}
    \caption{BB84 rate for every sublink across the path between Bob and Hadi.}
    \label{tab:dronebobtohadi}
\end{table}

The QKD rate of this scenario is thus limited by the drone-to-drone link and would be 0.079 bits per attempt for Bob and Hadi. Note that this gives us only a first estimate of the BB84 rate between two Qlients in Quantum Cities linked with high-altitude balloons. It shows however the feasibility of such free-space links, which could come as a more accessible solution to perform quantum communication when satellites are not available. Theoretically, two balloons at 10~km altitude can be separated by a maximum 714~km and still be visible in the horizon. We leave a more detailed study of this kind of high-altitude balloon links for future work.

\subsection{Towards quantum Internet applications}

\begin{figure}[!ht]
    \centering
    \includegraphics[width=12cm]{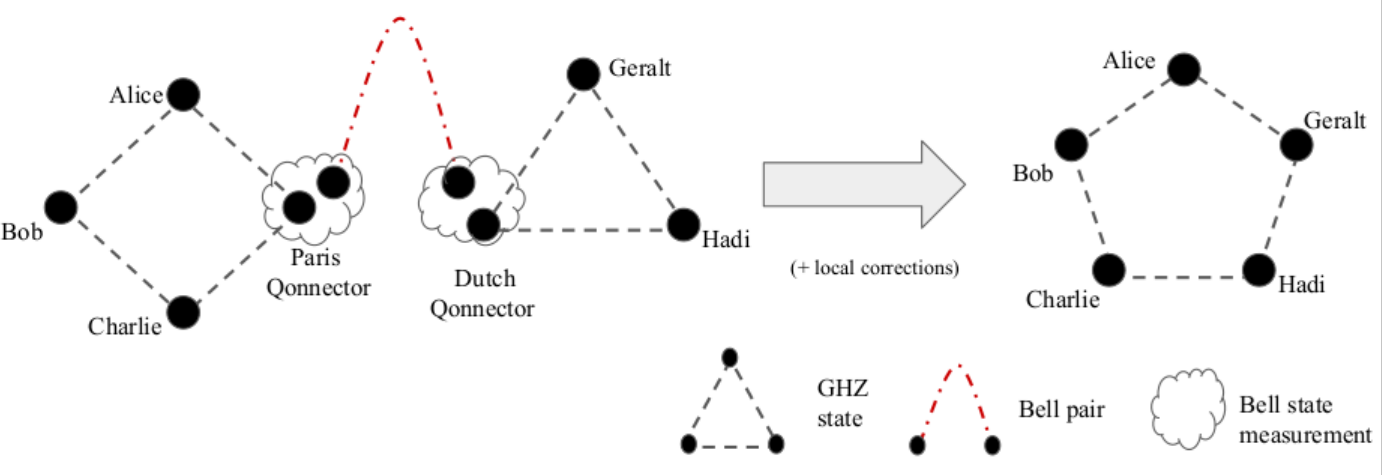}
    \caption{By making Bell state measurements and local operations, two GHZ states generated in Paris and the Netherlands can be transformed into one single shared GHZ state, consuming one Bell pair shared between the two metropolitan networks.}
    \label{fig:graphman}
\end{figure}

Once entanglement is generated between the Qonnectors of our two Quantum cities, several applications, beyond quantum key distribution, becomes available to the Qlients. For instance, as we detailed in~\cite{QCity}, sharing a GHZ state, $\frac{\ket{0}^{\bigotimes n}+\ket{1}^{\bigotimes n}}{\sqrt{2}}$ ~\cite{GHZ}, to the Qlients and processing the measurement outcomes enables conference key agreement protocols~\cite{CKAPappa,Murta_2020}, the multipartite counterpart of QKD allowing $n$ parties to get a secure shared key, or anonymous transmission ~\cite{Anonymity} and electronic voting ~\cite{fedeVoting} with high security and privacy guarantees. Through Bell state measurements and local operations, two GHZ states and a Bell pair can be transformed into a bigger GHZ state as we show in Fig.~\ref{fig:graphman}. For more information about graph state manipulation, see e.g.~\cite{GraphStateAxel,Meignant_2019}. These techniques are interesting for scaling up quantum networking applications, including for satellite quantum communication in the long run.

The implementation of protocols relying on multipartite entanglement in a realistic adversarial framework typically relies on performing multiple rounds of state verification~\cite{MEVresistant} in between rounds of actual use of the state. For this, high-fidelity states need to be generated and routed efficiently through the network. This is quite challenging with present technology. As explained in detail in~\cite{QCity}, photonic GHZ states are created by simultaneously creating Bell pairs by spontaneous parametric down conversion that are then entangled using fusion gates~\cite{EntanglementFusionXP}. All these processes are probabilistic and this makes the probability of succeeding in creating a GHZ state quite low: on the order of $10^{-3}$ for 3 or 4 qubits GHZ states and $10^{-5}$ for 5 or 6 qubits GHZ state. State generation using single-photon sources may eventually improve these rates.\\

Our Qloud architecture is compatible with the deployment of such applications that will become available as technology progresses, and with other ones yet to be discovered that will rely on equivalent resources. Considering upgradability to accommodate advanced applications in the design of near-term quantum networks is important such that the developed technology can be efficiently used at all network stages. We also remark that the Qloud architecture is convenient for new users to join as they only has to connect to their Qonnector in a Quantum City. Future work will explore how communication between bacQbone nodes could be optimized to facilitate communication of two Qlients in distant Quantum Cities, for example Dina and Alice in Fig~\ref{fig:Qloud}.


\section{Conclusion}
In this work, we have studied the feasibility of satellite quantum communication between two Quantum Cities in Europe through the development and use of a simulation library for the NetSquid quantum-network simulator. Through these simulations, we were able to perform parameter exploration of quantum communication from satellite to ground stations, using real satellite data, by computing the transmissivity for each individual photon. Our code is available on GitHub~\cite{github,githubMatteo} and is modular. We then embedded this analysis in a quantum network setting, to explore the relevance of satellite communication in concrete scenarios. We showed the performance of two different QKD configurations in a specific realistic setting linking two European cities. The underlying Qloud architecture shown in Fig.~\ref{fig:Qloud} minimizes end user hardware while facilitating routing of entanglement, and also allowing for several alternatives for creating correlation between the users. We also discussed alternatives to satellite communication for connecting different Quantum Cities, such as high-altitude balloons. Combinations of terrestrial, free-space and satellite links will likely be required for the realization of the full-scale Quantum Internet. We believe that simulation tools like the ones developed for this work are instrumental for assessing the necessity and gain of deploying such resource-intensive infrastructures, considering all relevant trade-offs in performance, service availability, security, etc., for promising quantum networking use cases.



\section{Acknowledgements}
This project has received funding from the European Union's Horizon Europe research and innovation program under the grant agreements No 101102140 (QIA) and 101114043 (QSNP), and under the project QUANGO (grant no. 101004341). We acknowledge support from the French National Research Agency (ANR) through the project SoLuQS as well as the Dutch National Growth Fund, as part of the Quantum Delta NL programme. This work was also supported by the JST Moonshot R\&D program under Grants JPMJMS226C

\bibliographystyle{unsrt}
\bibliography{main}

\begin{thebibliography}{10}

\bibitem{QIavision}
Stephanie Wehner, David Elkouss, and Ronald Hanson.
\newblock Quantum internet: A vision for the road ahead.
\newblock {\em Science}, 362(6412), 2018.

\bibitem{SecurityQKD}
Valerio Scarani, Helle Bechmann-Pasquinucci, Nicolas~J. Cerf, Miloslav
  Du\ifmmode~\check{s}\else \v{s}\fi{}ek, Norbert L\"utkenhaus, and Momtchil
  Peev.
\newblock The security of practical quantum key distribution.
\newblock {\em Rev. Mod. Phys.}, 81:1301--1350, Sep 2009.

\bibitem{Leaderelection}
Maor Ganz.
\newblock Quantum leader election.
\newblock {\em Quantum Information Processing}, 16, 10 2009.

\bibitem{SecretSharing}
Mark Hillery, Vladimír Bužek, and André Berthiaume.
\newblock Quantum secret sharing.
\newblock {\em Physical Review A}, 59(3):1829–1834, Mar 1999.

\bibitem{RefCommComplexity}
Gilles Brassard.
\newblock Quantum communication complexity.
\newblock {\em Foundations of Physics}, 33(11):1593–1616, November 2003.

\bibitem{delegated1}
Joseph Fitzsimons.
\newblock Private quantum computation: An introduction to blind quantum
  computing and related protocols.
\newblock {\em npj Quantum Information}, 3, 11 2016.

\bibitem{DelegatedQC}
Vedran Dunjko, Joseph~F. Fitzsimons, Christopher Portmann, and Renato Renner.
\newblock Composable security of delegated quantum computation.
\newblock In Palash Sarkar and Tetsu Iwata, editors, {\em Advances in
  Cryptology -- ASIACRYPT 2014}, pages 406--425, Berlin, Heidelberg, 2014.
  Springer Berlin Heidelberg.

\bibitem{DANOS200773}
Vincent Danos, Ellie D'Hondt, Elham Kashefi, and Prakash Panangaden.
\newblock Distributed measurement-based quantum computation.
\newblock {\em Electronic Notes in Theoretical Computer Science}, 170:73--94,
  2007.
\newblock Proceedings of the 3rd International Workshop on Quantum Programming
  Languages (QPL 2005).

\bibitem{shettell2022private}
Nathan Shettell, Majid Hassani, and Damian Markham.
\newblock Private network parameter estimation with quantum sensors, 2022.

\bibitem{MEVresistant}
Anna Pappa, André Chailloux, Stephanie Wehner, Eleni Diamanti, and Iordanis
  Kerenidis.
\newblock Multipartite entanglement verification resistant against dishonest
  parties.
\newblock {\em Physical Review Letters}, 108, 12 2011.

\bibitem{Anonymity}
Anupama Unnikrishnan, Ian MacFarlane, Richard Yi, Eleni Diamanti, Damian
  Markham, and Iordanis Kerenidis.
\newblock Anonymity for practical quantum networks.
\newblock {\em Physical Review Letters}, 122, 11 2018.

\bibitem{fedeVoting}
Federico Centrone, Eleni Diamanti, and Iordanis Kerenidis.
\newblock Quantum protocol for electronic voting without election authorities.
\newblock {\em Phys. Rev. Applied}, 18:014005, 2021.

\bibitem{ConferenceKeyAgreement}
Gláucia Murta, Federico Grasselli, Hermann Kampermann, and Dagmar Bruß.
\newblock Quantum conference key agreement: A review.
\newblock {\em Adv. Quantum Technol.}, 3:2000025, 2020.

\bibitem{Meignant_2019}
Cl{\'{e}}ment Meignant, Damian Markham, and Fr{\'{e}}d{\'{e}}ric Grosshans.
\newblock Distributing graph states over arbitrary quantum networks.
\newblock {\em Physical Review A}, 100(5), nov 2019.

\bibitem{Bughalo_2023}
Luis Bugalho, Bruno~C. Coutinho, Francisco~A. Monteiro, and Yasser Omar.
\newblock Distributing multipartite entanglement over noisy quantum networks.
\newblock {\em Quantum}, 7:920, nov 2023.

\bibitem{QIRG}
IETF Quantum Internet~Research Group.
\newblock Architectural principles for a quantum internet,
  https://datatracker.ietf.org/doc/draft-irtf-qirg-principles/.

\bibitem{BristolQCity}
Siddarth~Koduru Joshi, Djeylan Aktas, Sören Wengerowsky, Martin Lončarić,
  Sebastian~Philipp Neumann, Bo~Liu, Thomas Scheidl, Guillermo~Currás Lorenzo,
  Željko Samec, Laurent Kling, and et~al.
\newblock A trusted node–free eight-user metropolitan quantum communication
  network.
\newblock {\em Science Advances}, 6(36):eaba0959, Sep 2020.

\bibitem{ChinaQKDNetwork}
Yu-Ao Chen, Qiang Zhang, Teng-Yun Chen, Wen-Qi Cai, Sheng-Kai Liao, Jun Zhang,
  Kai Chen, Juan Yin, Ji-Gang Ren, Zhu Chen, Sheng-Long Han, Qing Yu, Ken
  Liang, Fei Zhou, Xiao Yuan, Mei-Sheng Zhao, Tian-Yin Wang, Xiao Jiang, Liang
  Zhang, Wei-Yue Liu, Yang Li, Qi~Shen, Yuan Cao, Chao-Yang Lu, Rong Shu,
  Jian-Yu Wang, Li~Li, Nai-Le Liu, Feihu Xu, Xiang-Bin Wang, Cheng-Zhi Peng,
  and Jian-Wei Pan.
\newblock {An integrated space-to-ground quantum communication network over
  4,600 kilometres}.
\newblock {\em Nature}, 589(7841):214--219, 2021.

\bibitem{PLOB}
Stefano Pirandola, Riccardo Laurenza, Carlo Ottaviani, and Leonardo Banchi.
\newblock Fundamental limits of repeaterless quantum communications.
\newblock {\em Nature Communications}, 8(1), apr 2017.

\bibitem{Takeoka_2014}
Masahiro Takeoka, Saikat Guha, and Mark~M. Wilde.
\newblock Fundamental rate-loss tradeoff for optical quantum key distribution.
\newblock {\em Nature Communications}, 5(1), oct 2014.

\bibitem{Repeater1}
H.-J. Briegel, W.~D\"ur, J.~I. Cirac, and P.~Zoller.
\newblock Quantum repeaters: The role of imperfect local operations in quantum
  communication.
\newblock {\em Phys. Rev. Lett.}, 81:5932--5935, Dec 1998.

\bibitem{Repeater2}
William~J. Munro, Koji Azuma, Kiyoshi Tamaki, and Kae Nemoto.
\newblock Inside quantum repeaters.
\newblock {\em IEEE Journal of Selected Topics in Quantum Electronics},
  21(3):78--90, 2015.

\bibitem{sangouard2011quantum}
Nicolas Sangouard, Christoph Simon, Hugues de~Riedmatten, and Nicolas Gisin.
\newblock Quantum repeaters based on atomic ensembles and linear optics.
\newblock {\em Rev. Mod. Phys.}, 83:33--80, Mar 2011.

\bibitem{azuma2022quantum}
Koji Azuma, Sophia~E Economou, David Elkouss, Paul Hilaire, Liang Jiang,
  Hoi-Kwong Lo, and Ilan Tzitrin.
\newblock Quantum repeaters: From quantum networks to the quantum internet.
\newblock {\em arXiv preprint arXiv:2212.10820}, 2022.

\bibitem{Bhaskar_2020}
M.~K. Bhaskar, R.~Riedinger, B.~Machielse, D.~S. Levonian, C.~T. Nguyen, E.~N.
  Knall, H.~Park, D.~Englund, M.~Lon{\v{c}}ar, D.~D. Sukachev, and M.~D. Lukin.
\newblock Experimental demonstration of memory-enhanced quantum communication.
\newblock {\em Nature}, 580(7801):60--64, mar 2020.

\bibitem{Repeater3}
M.~Pompili, S.~L.~N. Hermans, S.~Baier, H.~K.~C. Beukers, P.~C. Humphreys,
  R.~N. Schouten, R.~F.~L. Vermeulen, M.~J. Tiggelman, L.~dos Santos~Martins,
  B.~Dirkse, S.~Wehner, and R.~Hanson.
\newblock Realization of a multinode quantum network of remote solid-state
  qubits.
\newblock {\em Science}, 372(6539):259--264, 2021.

\bibitem{PompiliScience2021}
M.~Pompili, S.~L.~N. Hermans, S.~Baier, H.~K.~C. Beukers, P.~C. Humphreys,
  R.~N. Schouten, R.~F.~L. Vermeulen, M.~J. Tiggelman, L.~dos Santos~Martins,
  B.~Dirkse, S.~Wehner, and R.~Hanson.
\newblock Realization of a multinode quantum network of remote solid-state
  qubits.
\newblock {\em Science}, 372(6539):259--264, 2021.

\bibitem{Lago_Rivera_2021}
Dario Lago-Rivera, Samuele Grandi, Jelena~V. Rakonjac, Alessandro Seri, and
  Hugues de~Riedmatten.
\newblock Telecom-heralded entanglement between multimode solid-state quantum
  memories.
\newblock {\em Nature}, 594(7861):37--40, jun 2021.

\bibitem{repeaterNV}
Filip Rozpedek, Raja Yehia, Kenneth Goodenough, Maximilian Ruf, Peter~C.
  Humphreys, Ronald Hanson, Stephanie Wehner, and David Elkouss.
\newblock Near-term quantum-repeater experiments with nitrogen-vacancy centers:
  Overcoming the limitations of direct transmission.
\newblock {\em Phys. Rev. A}, 99:052330, May 2019.

\bibitem{Ruf_2021}
Maximilian Ruf, Noel~H. Wan, Hyeongrak Choi, Dirk Englund, and Ronald Hanson.
\newblock Quantum networks based on color centers in diamond.
\newblock {\em Journal of Applied Physics}, 130(7):070901, aug 2021.

\bibitem{avis2022requirements}
Guus Avis, Francisco~Ferreira da~Silva, Tim Coopmans, Axel Dahlberg, Hana
  Jirovsk{\'a}, David Maier, Julian Rabbie, Ariana Torres-Knoop, and Stephanie
  Wehner.
\newblock Requirements for a processing-node quantum repeater on a real-world
  fiber grid.
\newblock {\em arXiv preprint arXiv:2207.10579}, 2022.

\bibitem{Bonato2009}
C~Bonato, A~Tomaello, V~Da Deppo, G~Naletto, and P~Villoresi.
\newblock Feasibility of satellite quantum key distribution.
\newblock {\em New Journal of Physics}, 11(4):045017, April 2009.

\bibitem{Bourgoin2013}
J-P Bourgoin, E~Meyer-Scott, B~L Higgins, B~Helou, C~Erven, H~H\"{u}bel,
  B~Kumar, D~Hudson, I~D{\textquotesingle}Souza, R~Girard, R~Laflamme, and
  T~Jennewein.
\newblock A comprehensive design and performance analysis of low earth orbit
  satellite quantum communication.
\newblock {\em New Journal of Physics}, 15(2):023006, February 2013.

\bibitem{SchmittManderbach2007}
Tobias Schmitt-Manderbach, Henning Weier, Martin F\"{u}rst, Rupert Ursin, Felix
  Tiefenbacher, Thomas Scheidl, Josep Perdigues, Zoran Sodnik, Christian
  Kurtsiefer, John~G. Rarity, Anton Zeilinger, and Harald Weinfurter.
\newblock Experimental demonstration of free-space decoy-state quantum key
  distribution over 144~km.
\newblock {\em Physical Review Letters}, 98(1), January 2007.

\bibitem{Nauerth2013}
Sebastian Nauerth, Florian Moll, Markus Rau, Christian Fuchs, Joachim Horwath,
  Stefan Frick, and Harald Weinfurter.
\newblock Air-to-ground quantum communication.
\newblock {\em Nature Photonics}, 7(5):382--386, March 2013.

\bibitem{Wang2013}
Jian-Yu Wang, Bin Yang, Sheng-Kai Liao, Liang Zhang, Qi~Shen, Xiao-Fang Hu,
  Jin-Cai Wu, Shi-Ji Yang, Hao Jiang, Yan-Lin Tang, Bo~Zhong, Hao Liang,
  Wei-Yue Liu, Yi-Hua Hu, Yong-Mei Huang, Bo~Qi, Ji-Gang Ren, Ge-Sheng Pan,
  Juan Yin, Jian-Jun Jia, Yu-Ao Chen, Kai Chen, Cheng-Zhi Peng, and Jian-Wei
  Pan.
\newblock Direct and full-scale experimental verifications towards
  ground{\textendash}satellite quantum key distribution.
\newblock {\em Nature Photonics}, 7(5):387--393, April 2013.

\bibitem{Vallone2015}
Giuseppe Vallone, Davide Bacco, Daniele Dequal, Simone Gaiarin, Vincenza
  Luceri, Giuseppe Bianco, and Paolo Villoresi.
\newblock Experimental satellite quantum communications.
\newblock {\em Physical Review Letters}, 115(4), July 2015.

\bibitem{Gnthner2017}
Kevin G\"{u}nthner, Imran Khan, Dominique Elser, Birgit Stiller, \"{O}mer
  Bayraktar, Christian~R. M\"{u}ller, Karen Saucke, Daniel Tr\"{o}ndle, Frank
  Heine, Stefan Seel, Peter Greulich, Herwig Zech, Bj\"{o}rn G\"{u}tlich,
  Sabine Philipp-May, Christoph Marquardt, and Gerd Leuchs.
\newblock Quantum-limited measurements of optical signals from a geostationary
  satellite.
\newblock {\em Optica}, 4(6):611, June 2017.

\bibitem{Lu2022}
Chao-Yang Lu, Yuan Cao, Cheng-Zhi Peng, and Jian-Wei Pan.
\newblock Micius quantum experiments in space.
\newblock {\em Reviews of Modern Physics}, 94(3), July 2022.

\bibitem{Liao2017}
Sheng-Kai Liao, Wen-Qi Cai, Wei-Yue Liu, Liang Zhang, Yang Li, Ji-Gang Ren,
  Juan Yin, Qi~Shen, Yuan Cao, Zheng-Ping Li, Feng-Zhi Li, Xia-Wei Chen, Li-Hua
  Sun, Jian-Jun Jia, Jin-Cai Wu, Xiao-Jun Jiang, Jian-Feng Wang, Yong-Mei
  Huang, Qiang Wang, Yi-Lin Zhou, Lei Deng, Tao Xi, Lu~Ma, Tai Hu, Qiang Zhang,
  Yu-Ao Chen, Nai-Le Liu, Xiang-Bin Wang, Zhen-Cai Zhu, Chao-Yang Lu, Rong Shu,
  Cheng-Zhi Peng, Jian-Yu Wang, and Jian-Wei Pan.
\newblock Satellite-to-ground quantum key distribution.
\newblock {\em Nature}, 549(7670):43--47, August 2017.

\bibitem{Yin2020}
Juan Yin, Yu-Huai Li, Sheng-Kai Liao, Meng Yang, Yuan Cao, Liang Zhang, Ji-Gang
  Ren, Wen-Qi Cai, Wei-Yue Liu, Shuang-Lin Li, Rong Shu, Yong-Mei Huang, Lei
  Deng, Li~Li, Qiang Zhang, Nai-Le Liu, Yu-Ao Chen, Chao-Yang Lu, Xiang-Bin
  Wang, Feihu Xu, Jian-Yu Wang, Cheng-Zhi Peng, Artur~K. Ekert, and Jian-Wei
  Pan.
\newblock Entanglement-based secure quantum cryptography over 1, 120
  kilometres.
\newblock {\em Nature}, 582(7813):501--505, June 2020.

\bibitem{Ren2017}
Ji-Gang Ren, Ping Xu, Hai-Lin Yong, Liang Zhang, Sheng-Kai Liao, Juan Yin,
  Wei-Yue Liu, Wen-Qi Cai, Meng Yang, Li~Li, Kui-Xing Yang, Xuan Han,
  Yong-Qiang Yao, Ji~Li, Hai-Yan Wu, Song Wan, Lei Liu, Ding-Quan Liu, Yao-Wu
  Kuang, Zhi-Ping He, Peng Shang, Cheng Guo, Ru-Hua Zheng, Kai Tian, Zhen-Cai
  Zhu, Nai-Le Liu, Chao-Yang Lu, Rong Shu, Yu-Ao Chen, Cheng-Zhi Peng, Jian-Yu
  Wang, and Jian-Wei Pan.
\newblock Ground-to-satellite quantum teleportation.
\newblock {\em Nature}, 549(7670):70--73, August 2017.

\bibitem{Liorni2021}
Carlo Liorni, Hermann Kampermann, and Dagmar Bru{\ss}.
\newblock Quantum repeaters in space.
\newblock {\em New Journal of Physics}, 23(5):053021, may 2021.

\bibitem{Khatri2021}
Sumeet Khatri, Anthony~J. Brady, Ren{\'{e}}e~A. Desporte, Manon~P. Bart, and
  Jonathan~P. Dowling.
\newblock Spooky action at a global distance: analysis of space-based
  entanglement distribution for the quantum internet.
\newblock {\em npj Quantum Information}, 7(1), jan 2021.

\bibitem{Boone2015}
K.~Boone, J.-P. Bourgoin, E.~Meyer-Scott, K.~Heshami, T.~Jennewein, and
  C.~Simon.
\newblock Entanglement over global distances via quantum repeaters with
  satellite links.
\newblock {\em Physical Review A}, 91(5), may 2015.

\bibitem{coopmans2021netsquid}
Tim Coopmans, Robert Knegjens, Axel Dahlberg, David Maier, Loek Nijsten, Julio
  de~Oliveira~Filho, Martijn Papendrecht, Julian Rabbie, Filip Rozp{\k{e}}dek,
  Matthew Skrzypczyk, et~al.
\newblock Netsquid, a network simulator for quantum information using discrete
  events.
\newblock {\em Communications Physics}, 4(1):1--15, 2021.

\bibitem{Netsquid}
Netsquid website, https://netsquid.org/.

\bibitem{QCity}
Raja Yehia, Simon Neves, Eleni Diamanti, and Iordanis Kerenidis.
\newblock Quantum city: simulation of a practical near-term metropolitan
  quantum network.
\newblock {\em arXiv preprint arXiv:2211.01190}, 2022.

\bibitem{github}
Raja Yehia.
\newblock Github repository for the netsquid simulation modules,
  https://github.com/rajayehia/quantumcity.

\bibitem{githubMatteo}
Matteo Schiavon and Tim Coopmans.
\newblock Netsquid-freespace,
  https://github.com/matteoschiav/netsquid-freespace.

\bibitem{Moll2019}
Florian Moll, Thierry Botter, Christoph Marquardt, David Pusey, Amita Shrestha,
  Andrew Reeves, Kevin Jaksch, Kevin Gunthner, \"{O}mer Bayraktar, Christian
  Mueller-Hirschkorn, Alberto~D. Gallardo, Dionisio~Diaz Gonzalez, Wenjamin
  Rosenfeld, Peter Freiwang, Gerd Leuchs, and Harald Weinfurter.
\newblock Stratospheric {QKD}: feasibility analysis and free-space optics
  system concept.
\newblock In Mark~T. Gruneisen, Miloslav Dusek, John~G. Rarity, and Paul~M.
  Alsing, editors, {\em Quantum Technologies and Quantum Information Science
  V}. {SPIE}, October 2019.

\bibitem{Vu2020}
Minh~Quang Vu, Thanh~V. Pham, Ngoc~T. Dang, and Anh~T. Pham.
\newblock Design and performance of relay-assisted satellite free-space optical
  quantum key distribution systems.
\newblock {\em {IEEE} Access}, 8:122498--122510, 2020.

\bibitem{Chu2021}
Yi~Chu, Ross Donaldson, Rupesh Kumar, and David Grace.
\newblock Feasibility of quantum key distribution from high altitude platforms.
\newblock {\em Quantum Science and Technology}, 6(3):035009, June 2021.

\bibitem{PhotonicIntegrationRoadmap}
Galan Moody and et~al.
\newblock 2022 roadmap on integrated quantum photonics.
\newblock {\em J. Phys: Photonics}, 4:012501, 2022.

\bibitem{Kneizys1988}
F.~X. Kneizys, E.~P. Shettle, L.~W. Abreu, J.~H. Chetwynd, and G.~P. Anderson.
\newblock {\em Users Guide to LOWTRAN 7}.
\newblock Air Force Geophysics Laboratory, 1988.

\bibitem{kneizys1978atmospheric}
Francis~X Kneizys.
\newblock Atmospheric transmittance and radiance: The lowtran code.
\newblock In {\em Optical Properties of the Atmosphere}, volume 142, pages
  6--8. International Society for Optics and Photonics, 1978.

\bibitem{SatelliteModel}
D.~Yu. Vasylyev, A.~A. Semenov, and W.~Vogel.
\newblock Toward global quantum communication: Beam wandering preserves
  nonclassicality.
\newblock {\em Phys. Rev. Lett.}, 108:220501, Jun 2012.

\bibitem{LuisVictorFeasability}
Daniele Dequal, Luis Trigo~Vidarte, Victor Roman~Rodriguez, Giuseppe Vallone,
  Paolo Villoresi, Anthony Leverrier, and Eleni Diamanti.
\newblock Feasibility of satellite-to-ground continuous-variable quantum key
  distribution.
\newblock {\em npj Quantum Information}, 7(1), Jan 2021.

\bibitem{SatelliteModelComplex}
D.~Vasylyev, W.~Vogel, and F.~Moll.
\newblock Satellite-mediated quantum atmospheric links.
\newblock {\em Physical Review A}, 99(5), May 2019.

\bibitem{n2yo}
https://www.n2yo.com/.

\bibitem{orekit}
orekit library, https://www.orekit.org/.

\bibitem{MiciusSat}
Sheng-Kai Liao, Wen-Qi Cai, Johannes Handsteiner, Bo~Liu, Juan Yin, Liang
  Zhang, Dominik Rauch, Matthias Fink, Ji-Gang Ren, Wei-Yue Liu, Yang Li,
  Qi~Shen, Yuan Cao, Feng-Zhi Li, Jian-Feng Wang, Yong-Mei Huang, Lei Deng, Tao
  Xi, Lu~Ma, Tai Hu, Li~Li, Nai-Le Liu, Franz Koidl, Peiyuan Wang, Yu-Ao Chen,
  Xiang-Bin Wang, Michael Steindorfer, Georg Kirchner, Chao-Yang Lu, Rong Shu,
  Rupert Ursin, Thomas Scheidl, Cheng-Zhi Peng, Jian-Yu Wang, Anton Zeilinger,
  and Jian-Wei Pan.
\newblock Satellite-relayed intercontinental quantum network.
\newblock {\em Phys. Rev. Lett.}, 120:030501, Jan 2018.

\bibitem{BBM92}
Charles~H. Bennett, Gilles Brassard, and N.~David Mermin.
\newblock Quantum cryptography without bell's theorem.
\newblock {\em Phys. Rev. Lett.}, 68:557--559, Feb 1992.

\bibitem{couplingfiber}
Bernard Klein and John Degnan.
\newblock Optical antenna gain 1: Transmitting antennas.
\newblock {\em Applied optics}, 13:2134--41, 09 1974.

\bibitem{Valentinarticle}
Valentina~Marulanda Acosta, Daniele Dequal, Matteo Schiavon, Aurélie
  Montmerle-Bonnefois, Caroline~B. Lim, Jean-Marc Conan, and Eleni Diamanti.
\newblock Analysis of satellite-to-ground quantum key distribution with
  adaptive optics.
\newblock {\em arXiv preprint arXiv:2111.06747}, 2021.

\bibitem{SatConstellation}
Sumeet Khatri, Anthony~J. Brady, Ren{\'{e}}e~A. Desporte, Manon~P. Bart, and
  Jonathan~P. Dowling.
\newblock Spooky action at a global distance: analysis of space-based
  entanglement distribution for the quantum internet.
\newblock {\em npj Quantum Information}, 7(1), jan 2021.

\bibitem{Multiplex2}
Dario Lago-Rivera, Samuele Grandi, Jelena~V. Rakonjac, Alessandro Seri, and
  Hugues de~Riedmatten.
\newblock Telecom-heralded entanglement between multimode solid-state quantum
  memories.
\newblock {\em Nature}, 594:030501, Jun 2021.

\bibitem{AllPhotonic}
Koji Azuma, Kiyoshi Tamaki, and Hoi-Kwong Lo.
\newblock All-photonic quantum repeaters.
\newblock {\em Nature Communications}, 6(1), apr 2015.

\bibitem{GHZ}
Daniel~M. Greenberger, Michael~A. Horne, and Anton Zeilinger.
\newblock {\em Going Beyond Bell's Theorem}, pages 69--72.
\newblock Springer Netherlands, Dordrecht, 1989.

\bibitem{CKAPappa}
Frederik Hahn, Jarn de~Jong, and Anna Pappa.
\newblock Anonymous quantum conference key agreement.
\newblock {\em PRX Quantum}, 1(2), Dec 2020.

\bibitem{Murta_2020}
Gláucia Murta, Federico Grasselli, Hermann Kampermann, and Dagmar Bruß.
\newblock Quantum conference key agreement: A review.
\newblock {\em Advanced Quantum Technologies}, 3(11):2000025, Sep 2020.

\bibitem{GraphStateAxel}
Axel Dahlberg, Jonas Helsen, and Stephanie Wehner.
\newblock How to transform graph states using single-qubit operations:
  computational complexity and algorithms.
\newblock {\em Quantum Science and Technology}, 5(4):045016, 2020.

\bibitem{EntanglementFusionXP}
Dik Bouwmeester, Jian-Wei Pan, Matthew Daniell, Harald Weinfurter, and Anton
  Zeilinger.
\newblock Observation of three-photon greenberger-horne-zeilinger entanglement.
\newblock {\em Physical Review Letters}, 82(7):1345, 1999.

\end{thebibliography}

\end{document}